\documentclass[a4paper,showpacs,twocolumn]{revtex4-1} 

\pdfoutput=1

\usepackage[pdftex]{graphicx}    
\usepackage{subfigure}
\usepackage[amssymb,textstyle]{SIunits}
\usepackage{amsmath,amssymb}
\usepackage{multirow}
\usepackage{lineno}
\usepackage{rotating}
\usepackage{slashed}
\usepackage{url}
\usepackage{hyperref}
\usepackage{xspace}
\usepackage{relsize}
\usepackage{microtype}
\usepackage{setspace}
\usepackage{bm}
\usepackage{booktabs}
\usepackage{xspace}
\usepackage{multirow}
\usepackage{bigstrut}
\usepackage{subfigure}
\usepackage{hyperref}
\usepackage{placeins}

\DeclareGraphicsExtensions{.pdf}

\hypersetup{
  colorlinks=true,
  linkcolor=red,
  citecolor=red,
  urlcolor=blue
}

\usepackage{preprintcover} 
\PreprintCoverPaperTitle{Measurement of  charged-particle event shape variables in inclusive $\sqrt{s}
  = \unit{7}{\tera\electronvolt}$ proton--proton interactions with the ATLAS detector}  
\PreprintIdNumber{CERN-PH-EP-2012-178} 
\PreprintCoverAbstract{

The measurement of charged-particle event shape variables is presented in 
inclusive inelastic $pp$ collisions at a center-of-mass energy 
of 7 TeV using the ATLAS detector at the LHC. 
The observables studied are the 
transverse thrust, thrust minor and transverse sphericity, each defined 
using the final-state charged particles' momentum components perpendicular to the beam direction.
Events with at least six charged particles are selected by a minimum-bias trigger.
In addition to the differential distributions, the evolution of each event shape variable as a function
of the leading charged particle transverse momentum, charged particle multiplicity and summed transverse momentum is presented. 
Predictions from several Monte Carlo models show significant deviations from data.

}  
\PreprintJournalName{PRD}  

\def\herwig{\textsc{Herwig}\xspace}
\def\pythia{\textsc{pythia}\xspace}
\DeclareRobustCommand{\pythiasix}{\pythia\!6\xspace}
\newcommand{\pythiaeight}{\pythia\!8\xspace}
\DeclareRobustCommand{\herwigpp}{\herwig\raisebox{0.2ex}{\smaller++}\xspace}

\newcommand{\ptlead}{\ensuremath{\pt^\text{lead}}\xspace}

\newcommand{\ptsum}{\ensuremath{\sum\mspace{-0.8mu}\pt}\xspace}
\newcommand{\etamod}{\ensuremath{|\eta\mspace{0.2mu}|}\xspace}
\newcommand{\Nchg}{\ensuremath{N_\text{ch}}\xspace}

\newcommand{\FigRef}[2][]{Fig#1.~\ref{#2}\xspace}
\newcommand{\SecRef}[2][]{Section#1~\ref{#2}\xspace}
\newcommand{\TabRef}[2][]{Table#1~\ref{#2}\xspace}

\def\mm{\milli\meter}
\newcommand{\abs}[1]{|#1|}
\def\pt{\ensuremath{p_{\mathrm{T}}}} 
\def\pT{\ensuremath{p_{\mathrm{T}}}} 
\def\TeV{\ifmmode {\mathrm{\ Te\kern -0.1em V}}\else
                   \textrm{Te\kern -0.1em V}\fi}%
\def\GeV{\ifmmode {\mathrm{\ Ge\kern -0.1em V}}\else
                   \textrm{Ge\kern -0.1em V}\fi}%
\def\MeV{\ifmmode {\mathrm{\ Me\kern -0.1em V}}\else
                   \textrm{Me\kern -0.1em V}\fi}%
\let\etaa=\eta
\def\eta{\ensuremath{\etaa}}%

\begin{document}


\title{Measurement of  charged-particle event shape variables in inclusive $\sqrt{s}
  = \unit{7}{\tera\electronvolt}$ proton--proton interactions with the ATLAS detector}

\author{The ATLAS Collaboration}


\begin{abstract}

The measurement of charged-particle event shape variables is presented in 
inclusive inelastic $pp$ collisions at a center-of-mass energy 
of 7 TeV using the ATLAS detector at the LHC. 
The observables studied are the 
transverse thrust, thrust minor and transverse sphericity, each defined 
using the final-state charged particles' momentum components perpendicular to the beam direction.
Events with at least six charged particles are selected by a minimum-bias trigger.
In addition to the differential distributions, the evolution of each event shape variable as a function
of the leading charged particle transverse momentum, charged particle multiplicity and summed transverse momentum is presented. 
Predictions from several Monte Carlo models show significant deviations from data.

\end{abstract}


\maketitle

\section{Introduction}

Event shape variables describe the structure of hadronic events and the properties of their energy flow. 
In this analysis, three event shape observables~\cite{Banfi:2004nk,Banfi:2010xy} are measured: the transverse 
thrust, the thrust minor and the transverse sphericity, each built from the momenta of charged particles using tracking information from proton-proton collisions at $\sqrt{s}=7$~TeV collected with the ATLAS 
detector~\cite{:2008zzm}. 
Event shape observables are among the simplest experimentally measured quantities and, depending on the  
events being considered, may have sensitivity to both the perturbative and non-perturbative
aspects of quantum chromodynamics (QCD).

Event shapes in hadronic collisions were investigated first at the Intersecting Storage Rings~\cite{Akesson1983354}
and at the Sp$\overline{\mathrm{p}}$S~\cite{arnison1987, Ansari:1987ig} at CERN to examine the 
emergence of jets, and later at Tevatron~\cite{Aaltonen:2011et}
to study the dependence of the event shape observables on the transverse energy of the leading jet 
and on contributions from the underlying event.
At the Large Hadron Collider (LHC), event shape observables were recently studied in inclusive interactions~\cite{Abelev:2012sk} and 
multi-jet events~\cite{Khachatryan:2011dx,Aad:2012np}. 
In $e^+e^-$ and $ep$ deep-inelastic scattering experiments,
the study of the energy flow in hadronic final states has allowed tests of the predictions of perturbative QCD, and the extraction of
a precise value for the strong coupling 
constant $\alpha_\text{S}$~\cite{MovillaFernandez:2001ed, Chekanov:2002xk, Achard:2002kv, Heister:2003aj, Abdallah:2004xe, 
Abbiendi:2004qz, Aktas:2005tz}.

The study of event shape observables in inclusive inelastic collisions plays
an important role in understanding soft-QCD 
processes at LHC center-of-mass energies~\cite{Wraight:2011ej},
where ``soft'' refers to interactions with low 
momentum transfer between the scattering particles.
Soft interactions cannot be reliably calculated from theory and are thus generally 
described by phenomenological models, usually implemented
in Monte Carlo (MC) event generators. These models contain many parameters whose
values are \textit{a priori} unknown and thus need to be constrained
by measurements.
Inclusive and semi-inclusive observables  sensitive to soft-QCD processes have been
measured at the LHC by the ATLAS~\cite{:2010ir, Aad:2010fh, Aad:2011qe}, 
CMS~\cite{Khachatryan:2010us, Chatrchyan:2011id} and ALICE~\cite{Aamodt:2010pp, ALICE:2011ac} collaborations.
The measurements presented in this paper can further
constrain the event generator models, which encapsulate our understanding of these soft processes.

In this analysis, the event shape observables are constructed from six or more primary charged particles 
in the pseudorapidity range $\etamod < 2.5$ and 
with transverse momentum
$\pt > \unit{0.5}{\GeV}$~\footnote{ATLAS uses a right-handed coordinate system with its origin at the nominal interaction point (IP) in the centre of the detector and the $z$-axis along the beam pipe. The $x$-axis points from the IP to the centre of the LHC ring, and the $y$-axis points upward. Cylindrical coordinates $(r,\phi)$ are used in the transverse plane, $\phi$ being the azimuthal angle around the beam pipe. The pseudorapidity is defined in terms of the polar angle $\theta$ as $\eta=-\ln\tan(\theta/2)$.}. 
Primary charged particles are defined as those with a mean proper lifetime 
$\tau > 30$~ps, 
produced either directly in the $pp$ interaction or from
the subsequent  decay of particles with a shorter lifetime. 
The particle level refers to particles as they emerge from the proton--proton interaction.  
The detector level corresponds to tracks as measured after interaction with the detector material, 
and includes the detector response.
The results are corrected for detector effects, using simulation, to obtain distributions of the
event shape variables defined at particle level which can be directly compared to MC models.

This paper is organized as follows: 
\SecRef{sec:es} defines the event shape variables;
the detector is described in \SecRef{sec:atlasdetector};
\SecRef{sec:mcmodels} discusses the MC models used in this analysis;
\SecRef{sec:selection} and \ref{sec:backgrounds} respectively describe the event selections and background contributions.
The correction of the data back to particle level, and estimation of the systematic uncertainties
are described in \SecRef{sec:corr} and \ref{sec:syst};
the results are discussed in \SecRef{sec:results} and finally the conclusions are presented in \SecRef{sec:conclusions}.

\section{Event Shape Observables}
\label{sec:es}

In particle collisions, event shape observables describe the geometric properties of the energy flow in the final state. 
A single event shape variable can distinguish in a continuous way between configurations in which all the particles are flowing 
(forward and backward) along a single axis and configurations where the energy is distributed uniformly over the $4\pi$ solid angle.
If defined as a ratio of measured quantities, the corresponding systematic uncertainties may be small.

In hadron collisions, where the center-of-mass frame of the 
interaction is usually boosted along the beam axis, event shape observables are
often defined in terms of the transverse momenta, 
which are Lorentz-invariant under such boosts.
Different formulations of event shape observables are possible; 
the most intuitive is to calculate the event shape from all particles in
an event. 
These are denoted by \textit{directly global} event shapes~\cite{Banfi:2004nk,Banfi:2010xy}.
In hadron collider experiments, it is not usually possible to detect
all particles in an event due to the finite detector acceptance, limited at 
small scattering angles by the presence of the beam pipe. 
Event shapes which include only particles from a restricted phase space 
in pseudorapidity $\eta$, are called \textit{central event shapes}: in this analysis 
charged particles within the range $|\eta|<2.5$ are used.
These central event shapes are nevertheless sensitive to non-perturbative effects at 
low momentum transfer and provide useful information about the event structure for development of models of proton--proton collisions.
The thrust is one of the most widely used event shape variables. 
The transverse thrust for a given event is defined as:
\begin{equation} \label{eq:thr}
 T_{\perp}= \max_{\hat{n}} \frac{\displaystyle\sum_{i} \abs{\vec{p}_{\text{T},i}\cdot\hat{n}}}{\displaystyle\sum_{i} \abs{\vec{p}_{\text{T},i}\,}}
\end{equation}
where the sum is performed over the transverse momenta
$\vec{p}_{\text{T,i\,}}$ of all charged particles in the event.
The thrust axis $\hat{n}_\text{T}$ is the unit vector $\hat{n}$ that maximizes the ratio in Eq.~(\ref{eq:thr}).
The transverse thrust ranges from $T_{\perp} = 1$  for a perfectly balanced, pencil-like, dijet topology 
to $T_{\perp} =  \langle \abs{\cos\psi} \rangle =  2/\pi $
for a circularly symmetric distribution of particles in the transverse plane, where $\psi$ is 
the azimuthal angle between the thrust axis and each respective particle.  
It is convenient to define the complement of $T_{\perp}$, $\tau_{\perp} = 1 - T_{\perp}$, 
to match the behavior of many event shape variables, which vanish in a balanced dijet topology.

The thrust axis $\hat{n}_\text{T}$ and the beam axis $\hat{z}$ define the \textit{event plane}. 
The transverse thrust minor measures the out-of-event-plane energy flow:
\begin{equation}
 T_{\text{M}}=\frac{\displaystyle\sum_{i} \abs{\vec{p}_{\text{T},i}\cdot
     \hat{n}_{\text{m}}}}{\displaystyle\sum_{i} \abs{\vec{p}_{\text{T},i}}}~, \quad \hat{n}_{\text{m}} = \hat{n}_{\text{T}} \times \hat{z}~\mbox{.}\nonumber
\end{equation}
The transverse thrust minor is $0$ for a pencil-like event in azimuth and $2/\pi$ for an isotropic event.

Another widely used event shape variable is the sphericity, $S$, which describes the event energy flow based on the momentum tensor,
\begin{equation}
	S^{\rm \alpha \beta} =  \frac{\displaystyle\sum_{i} p^{\rm \alpha}_{i} p^{\rm \beta}_{i}}{\displaystyle\sum_{i} \abs{\vec{p}_{i}}^{2}}~\mbox{,}\nonumber
\end{equation}
where the Greek indices represent the $x$, $y$, and $z$ components of the momentum of the particle $i$. The sphericity of the event is defined in terms of the two smallest eigenvalues of this tensor, $\lambda_2$ and $\lambda_3$:
\begin{equation}
	S=\frac{3}{2}({\lambda_2 + \lambda_3}).\nonumber
\end{equation}
The sphericity has values between 0 and 1, where a balanced dijet event corresponds to $S=0$ and an isotropic event to $S=1$. Sphericity is essentially a measure of the summed $\pT^2$ with respect to the event axis~\cite{Bjorken:1969wi,PhysRevLett35_1609}, where the event axis is defined as the line passing through the interaction point and oriented along the eigenvector associated with the largest eigenvalue, $\lambda_1$. 
Similarly to transverse thrust, the transverse sphericity, $S_{\perp}$, is defined in terms of the transverse components only: 
\begin{equation}
	S^{\,xy} = \sum_{i} \frac{1}{\abs{\vec{p}_{\text{T},i}}^{2}} \begin{bmatrix}
  p_{x,i}^2 & p_{x,i}\,p_{y,i} \\
  p_{x,i}\,p_{y,i} & p_{y,i}^2 \\
 \end{bmatrix}\nonumber
\end{equation}
and
\begin{equation}
	S_{\perp} = \frac{2\lambda^{xy}_{2}}{\lambda^{xy}_{1} + \lambda^{xy}_{2}}~\mbox{,}\nonumber
\end{equation}
where $\lambda^{xy}_{2} < \lambda^{xy}_{1}$ are the two eigenvalues of
$S^{xy}$.

The following distributions are measured:

\begin{itemize}

\item Normalized distributions: $(1/N_{\rm {ev}})dN_{\rm {ev}}/d\tau_\perp^{\rm {ch}}$, $(1/N_{\rm {ev}})dN_{\rm {ev}}/dT_M^{\rm {ch}}$, 
$(1/N_{\rm {ev}})dN_{\rm {ev}}/dS_\perp^{\rm {ch}}$;

\item Average values: $ \langle \tau_{\perp}^{\text{ch}} \rangle$, $ \langle T_{\text{M}}^{\text{ch}} \rangle $ and 
$\langle S_{\perp}^{\text{ch}} \rangle $ as functions of $N_{\text{ch}}$ and \ptsum ;

\end{itemize}
where $N_{\text{ev}}$ is the number of events with six or more charged particles within the selected kinematic range; $N_{\text{ch}}$ is the number of charged particles 
in an event; \ptsum is the scalar sum of the transverse momenta of the charged particles in the event. The event shape observables $\tau_{\perp}^{\text{ch}}$, $T_{\text{M}}^\text{{ch}}$ and $S_{\perp}^{\text{ch}}$ are defined as above, with the superscript indicating that they are
constructed from charged particles.%
The three normalized differential distributions are studied separately for:

\begin{itemize}
 \item $0.5~\mathrm{GeV} < \ptlead \leq 2.5~\mathrm{GeV}$
 \item $2.5~\mathrm{GeV} < \ptlead \leq 5.0~\mathrm{GeV}$
 \item $5.0~\mathrm{GeV} < \ptlead \leq 7.5~\mathrm{GeV}$
 \item $7.5~\mathrm{GeV} < \ptlead \leq 10.0~\mathrm{GeV}$
 \item $\ptlead > 10~\mathrm{GeV}$
\end{itemize} 
where \ptlead is the transverse momentum of the highest \pT\ (leading) charged particle.

\section{The ATLAS detector}
\label{sec:atlasdetector}

The ATLAS detector~\cite{:2008zzm} covers almost 
the full solid angle around the collision point with layers of tracking detectors, calorimeters 
and muon chambers. 
 The components that are relevant for this analysis are the tracking
detectors. 
The inner tracking detector has full coverage in azimuthal angle $\phi$ and 
covers the pseudorapidity range $\etamod~ < 2.5$.
It consists of a silicon pixel detector (pixel), a semiconductor tracker (SCT) and for $|\eta| < 2.0$,
a straw-tube transition radiation tracker (TRT).
These detectors, immersed in a 2~T axial magnetic field, are located at a 
radial distance from 
the beam line of
50.5--\unit{150}{\mm}, 299--\unit{560}{\mm} and 563--\unit{1066}{\mm},
respectively.
They provide position resolutions typically of \unit{10}{\micro\metre}, \unit{17}{\micro\metre} and
\unit{130}{\micro\metre} for the $r$--$\phi$ coordinate, and 
of \unit{115}{\micro\metre} and \unit{580}{\micro\metre} for 
the $z$ coordinate in the case of the pixel and SCT detectors.

The measurements presented here use events triggered by the minimum-bias trigger scintillator (MBTS) system~\cite{Aad:2012xs}.
The MBTS detectors are mounted at each end of the tracking detector 
at $z = \pm \unit{3.56}{\metre}$ and are segmented into eight sectors in azimuth and two concentric rings in pseudorapidity 
($2.09 < \etamod < 2.82$ and $2.82 < \etamod < 3.84$). 
The MBTS trigger was configured to require at least one hit above threshold from either side of the detector 
in coincidence with a fast beam-pickup device 
 ensuring that the event is compatible with a bunch crossing.


\section{Monte Carlo models}
\label{sec:mcmodels}

Monte Carlo (MC) event samples are used to
compute the detector acceptance and reconstruction efficiency,
determine background contributions, correct the measurements for
detector effects, and to calculate systematic uncertainties.
Finally, 
different phenomenological models implemented 
in the MC generators are compared to the data corrected to the particle level.

The \pythiasix~\cite{Sjostrand:2006za}, \pythiaeight~\cite{Sjostrand:2007gs} and 
\herwigpp~\cite{Bahr:2008pv,Gieseke:2011na} event generators were used to produce the simulated event samples 
for the analysis.
These generators implement leading-logarithm parton shower models matched to 
leading-order matrix element calculations with different
hadronization models and orderings for the parton shower.
The \pythiasix and \pythiaeight generators use a hadronization model based
upon fragmentation of color strings and a \pT-ordered or virtuality-ordered shower, 
whereas the \herwigpp generator implements a
cluster hadronization scheme 
with parton showering ordered
by emission angle.
The \pythiaeight generator uses a multi-parton interaction (MPI) 
model interleaved with both initial-state and final-state (ISR and FSR) radiation, and
all three processes compete against each other for
emission phase space in the resulting evolution.
The \herwigpp UE7-2 tune employs color reconnection.
Different settings of model parameters, tuned to reproduce the existing experimental data
were used for the MC generators. \TabRef{tab:mc} shows the different MC models
used in this paper.

\begin{table*}[ht]
  \singlespacing
  \caption{Details of the MC models used. It is emphasized that the tunes use data from different experiments
to constrain different processes, but for brevity only the \textit{data} which had the most
weight in each specific tune are shown. Here ``LHC'' indicates data taken at $\sqrt{s}=7$~TeV,
although $\sqrt{s}=900$~GeV data were also included in ATLAS tunes, with much smaller weight.
Some tunes are focused on describing the minimum-bias (MB) distributions better, while the rest
are tuned to describe the underlying event (UE) distributions, as indicated. Authors indicates a tune performed by the MC developers}
  \begin{center}
    \begin{tabular}{lllllll}
      \toprule
      Generator & Version & Tune & PDF &  Focus  & Data &  From  \\
      \midrule
       \pythiasix & 6.425  & AMBT1~\cite{atlastunesto2010data} & MRST LO**~\cite{Sherstnev:2007nd} & MB & Early LHC & ATLAS\\
       \pythiasix & 6.425 & AMBT2B~\cite{:mc11} & CTEQ6L1~\cite{Pumplin:2002vw}  & MB & LHC & ATLAS \\
       \pythiasix & 6.421 & DW~\cite{CDFtuneA} & CTEQ5L~\cite{Lai:1999wy} & UE  &  Tevatron  & CDF \\
       \pythiasix & 6.425 & Z1~\cite{tuneZ1} &  CTEQ5L & UE & LHC& CMS \\
       \pythiaeight & 8.157 & A2~\cite{:mc11nlo} & MSTW2008LO~\cite{Martin:2009iq} & MB & LHC & ATLAS   \\
       \herwigpp  &  2.5.1   & UE7-2~\cite{Gieseke:2011xy} & MRST LO** & UE & LHC & Authors  \\
       \herwigpp & 2.5.0 & Default & MRST LO**  & UE  & LHC & Authors   \\
       \bottomrule
    \end{tabular}
    \label{tab:mc}
  \end{center}
\end{table*}

The reference model for this analysis is chosen to be \pythiasix AMBT1. 
Samples generated with this tune were passed through
the ATLAS detector and trigger simulations~\cite{:2010wqa} based on GEANT4~\cite{Agostinelli:2002hh}
and then reconstructed and analyzed using the 
same procedure and software that are used for the data.
Reconstructed MC events are then used to correct the data for detector effects. 
The sample generated with an older version of \herwigpp, 2.5.0 with no additional tuning, was 
also passed through the full detector simulation and the analysis chain for systematic studies of unfolding corrections.

\section{Event and track selection}
\label{sec:selection}

The data used for the analysis presented here 
were collected  
in April 2010 
with a minimal prescale factor for the minimum-bias trigger.
The only further requirement for selecting the data sample
is that the MBTS trigger and all inner detector
subsystems were at nominal operating conditions.
In each event the reconstructed vertices are ordered by the $\sum\pt^2$
over the tracks assigned to each vertex,
and the vertex with the highest $\sum\pt^2$ is taken as the primary interaction vertex of the event.
To reduce the contribution from beam-related backgrounds and 
decays of long-lived particles,
and to minimize the systematic uncertainties, events are 
rejected if they contain any other vertex reconstructed with four or more tracks.

If there is only one vertex in the event, or
if any additional vertex in the event has three or fewer tracks, 
all tracks from the event that pass the track selection (described below) are retained.
After this selection, the fraction of events with more than one proton--proton 
interaction in the same bunch crossing (referred to as pile-up)
is found to be approximately $0.1\%$ and this residual contribution is therefore neglected. 
The average number of $pp$ interactions per bunch crossing during this
data-taking period was less than 0.15, indicating a negligible pile-up
contribution.
The MC samples used have no pile-up contribution.

Events are required to contain at least six
tracks that fulfill the following criteria:
  \begin{itemize}
  \item $\pt > \unit{0.5}{\GeV}$;
  \item $|\eta| < 2.5$;
  \item a minimum of one pixel and six SCT hits; 
  \item a hit in the innermost pixel layer, 
   if the corresponding pixel module was active;
  \item transverse and 
   longitudinal impact parameters with respect to the primary vertex,
  $|d_0| <\unit{1.5}{\mm}$ and $|z_0|\sin \theta < \unit{1.5}{\mm}$;
  \item a track-fit probability $\chi^2 
      > 0.01$ for tracks with $\pt > \unit{10}{\GeV}$ in order to remove mis-measured tracks.
   \end{itemize}

Tracks with $\pt > \unit{0.5}{\GeV}$ are less prone than
lower-\pt\ tracks to inefficiencies and systematic uncertainties resulting from
interactions with the material inside the tracking volume.

After event selection, the analysis is based on approximately 17 million events containing approximately 300 million tracks.
For the \pythiasix generator and for the \pythiaeight generator, which has a harder diffractive model than the former,  
the contribution 
to the event shape observables from diffractive events is negligible when requiring six or more tracks in the event.

\begin{figure}[htb]
 \begin{center}
   \includegraphics[width=.495\textwidth]{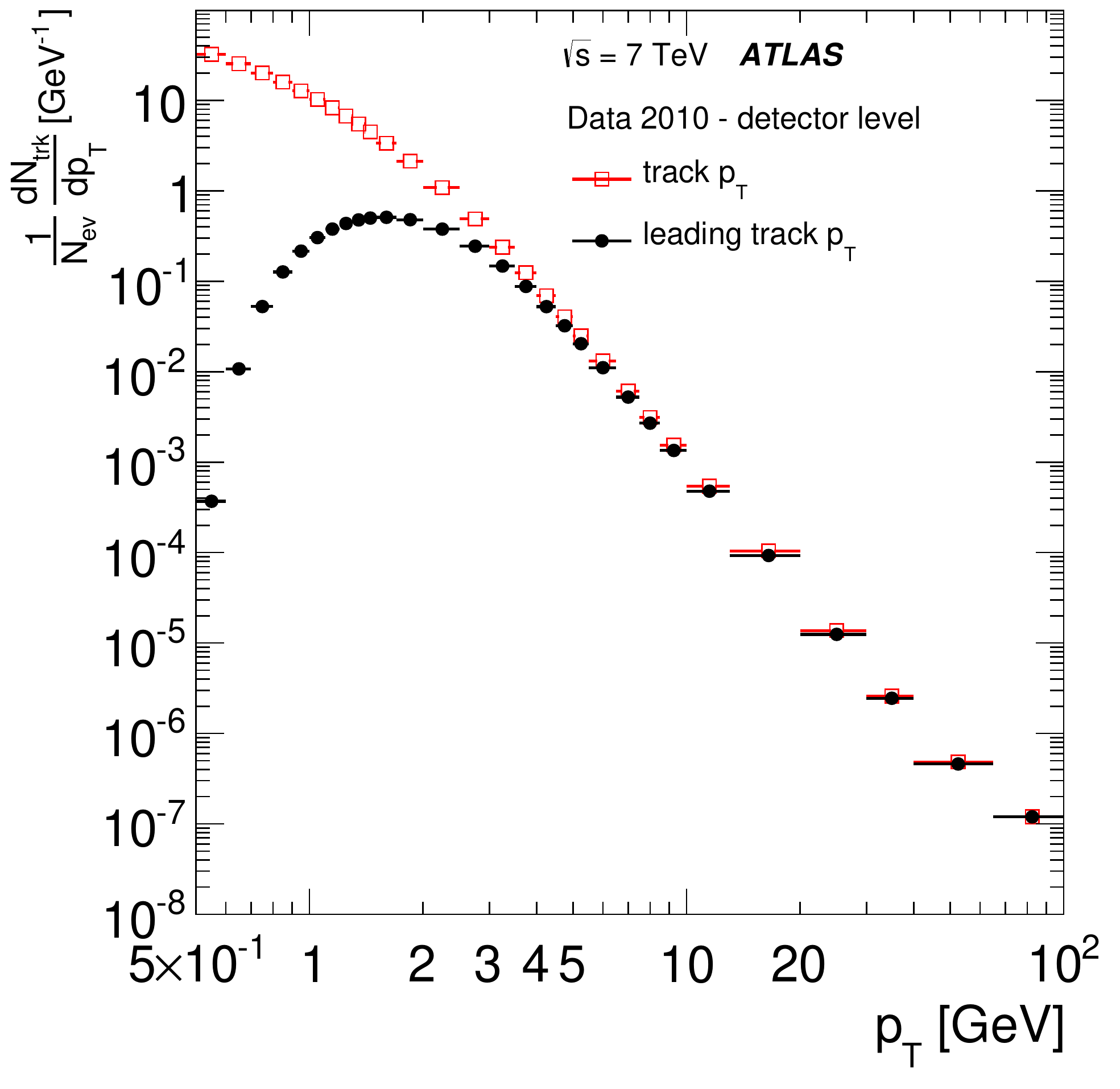}\hfill
    \end{center}
 \caption{The distribution of the transverse momentum of all tracks and of the leading transverse momentum track 
in data at detector level. The uncertainties shown are statistical. Where not visible, the statistical error is smaller than the marker size.}
 \label{fig:ptlead}
\end{figure}

The \pT\ distributions of all tracks and of the leading track in the selected event are shown in \FigRef{fig:ptlead}. 
The fraction of events in each \ptlead\ bin is shown in \TabRef{tab:noevents}.

\begin{table}
  \caption{Percentage of events in each \ptlead bin}
  \begin{center}
  \begin{tabular}{cc}
\ptlead bin [GeV] & Percentage of events\\ 
\hline 
$0.5 \textendash 2.5$ & $68.45$ \\ 
$2.5 \textendash 5.0$ & $28.20$ \\ 
$5.0 \textendash 7.5$ &  ~$2.65$\\ 
$7.5 \textendash 10.0$ & ~$0.47$ \\ 
$> 10.0$ & ~$0.23$ \\ 
\hline 
\end{tabular} 
  \label{tab:noevents}
  \end{center}
  \end{table}

\section{Background contributions}
\label{sec:backgrounds}

\subsection{Backgrounds}
Backgrounds comprise beam-induced events, due to beam-gas and beam-material interactions,
as well as non-beam backgrounds from cosmic-ray interactions and detector noise.
The contribution of these background events remaining after the event selection
is estimated using the number of pixel hits not associated with reconstructed tracks.
This multiplicity
includes unassigned hits from low-\pt\, looping tracks, but is dominated at
higher multiplicities by hits from charged particles produced in 
inelastic interactions of protons with the residual gas inside the beam pipe.
The vertex requirement removes most of the beam background events
and the residual contribution is below~0.1\%. As the level of background is very low, no
explicit background subtraction is performed.

\subsection{Secondary track fraction}
The primary charged particle multiplicities are measured from selected tracks
after correcting for the fractions of secondary and
poorly reconstructed tracks in the sample. The potential background from fake
tracks is found from MC studies to be less than 0.01\%~\cite{:2010ir}.

Non-primary tracks arise predominantly from hadronic interactions, photon
conversions to positron--electron pairs in the detector material and decays of long-lived particles.
For $\pt > \unit{0.5}{\GeV}$ the contribution from photon conversions is small.
The systematic uncertainty from secondary decays is included in the uncertainties 
associated with the tracking performance.

\section{Correction to particle level}
\label{sec:corr}

To facilitate comparison with theoretical predictions and other measurements,
the event shape distributions for charged particles are presented at particle level, after correction for trigger and event selection
efficiencies, as well as detector resolution effects.
A two-step correction procedure is used:
first, corrections for event selection efficiency are applied, followed by an
additional bin-by-bin correction to account for tracking inefficiencies, possible bin
migrations and any remaining detector effects.

\subsection{Event-level correction}
\label{sec:evcorr}

Trigger and vertexing efficiencies are taken from a previous analysis using the same data sample~\cite{:2010ir}. 
The efficiency of the MBTS trigger is determined from data using a control trigger 
and found to be fully efficient for the analysis requirement of at least six tracks. 
The vertex reconstruction efficiency is also measured in data by taking the ratio of the number of 
triggered events with a reconstructed vertex to the total number of triggered events. This ratio is also found to be very close to unity.
The total correction applied to account for events lost due to the trigger and vertex requirements is less than $1\%$  
and it varies very weakly with 
the number of tracks associated with the primary vertex.

\subsection{Bin-by-bin correction}
\label{sec:bbb}

The event shape observables presented here are 
sensitive to changes in the configuration of the selected tracks. 
Applying average track efficiencies to  individual tracks 
on a track-by-track basis and reweighting tracks distorts the event shape distribution. 
A more robust approach is to apply bin-by-bin corrections to find the event shape distribution at particle level.
Such a bin-by-bin correction is applied to all distributions after applying the event-level efficiency corrections
described above.

The correction factors $C_\text{bin}$ are evaluated separately in each
bin for each event shape observable,
\begin{equation*}
  C_\text{bin} = \frac{V^\text{Gen}_\text{bin}}{V^\text{Reco,~eff~corr}_\text{bin}},
\end{equation*}
where $V^\text{Gen}_\text{bin}$ and
$V^\text{Reco,~eff~corr}_\text{bin}$ represent the generator-level MC value of the bin content and the reconstructed MC value after applying
the event-level efficiency corrections for each bin, respectively. The corrected
value of the bin content for an observable is found by multiplying the measured bin content by
the corresponding correction factor.
The bin sizes are chosen to be consistent with the resolution of the correction procedure.

\begin{figure}
    
      \subfigure[]{
        \includegraphics[width=0.95\columnwidth]{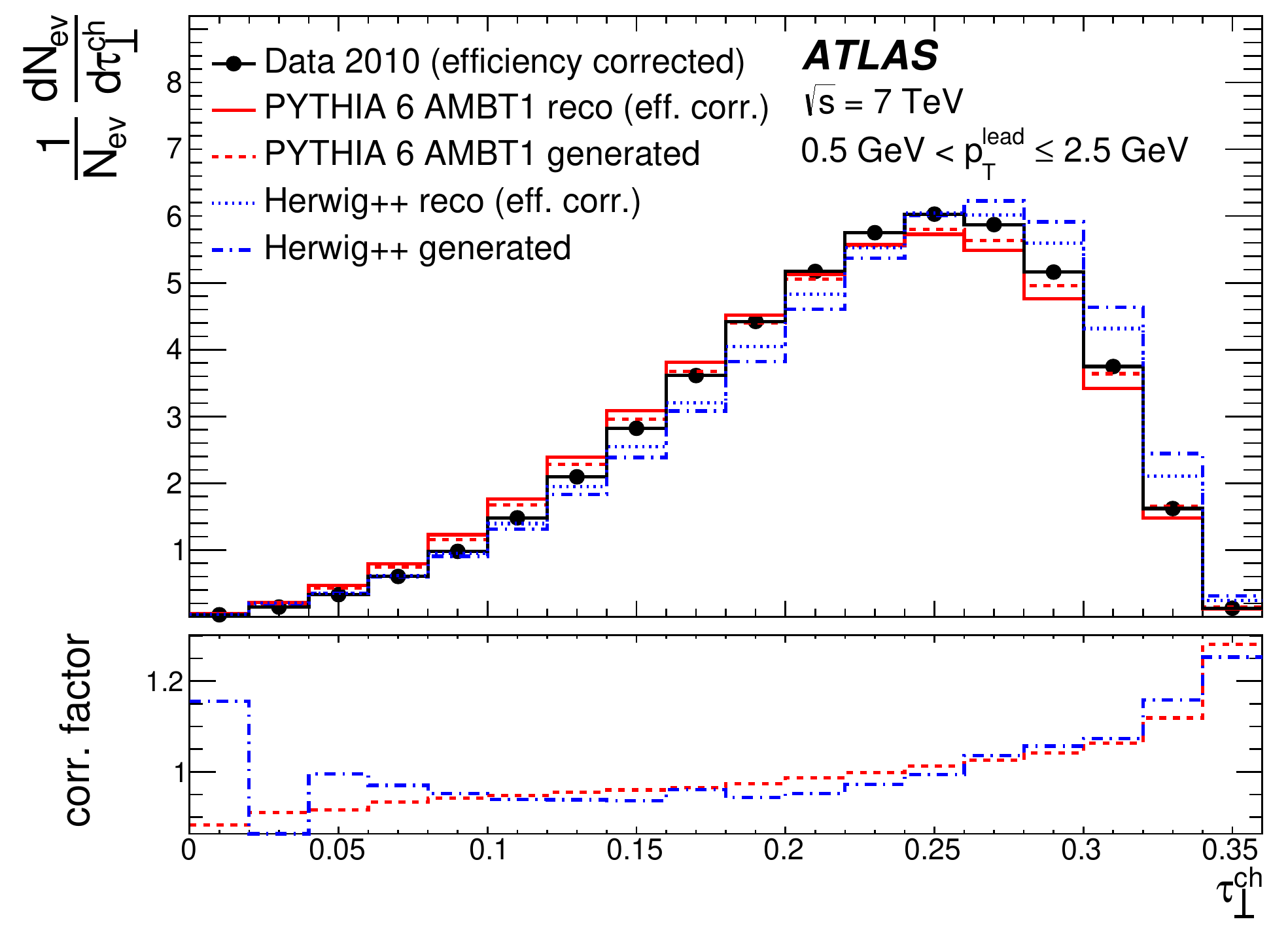}
        \label{fig:U-thr}}
      \subfigure[]{
        \includegraphics[width=0.95\columnwidth]{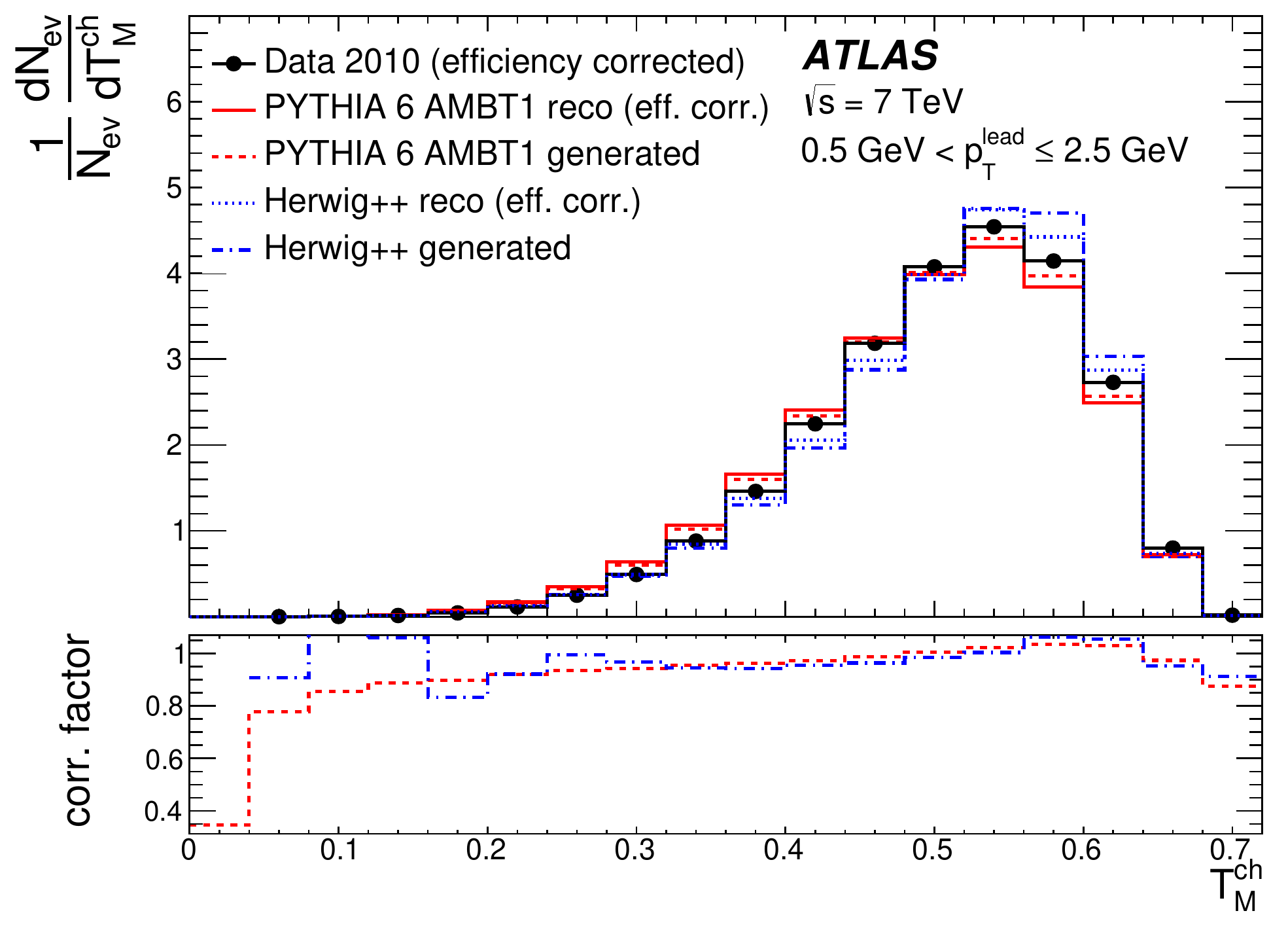}
        \label{fig:U-thrm}}
    
	  \subfigure[]{
	    \includegraphics[width=0.95\columnwidth]{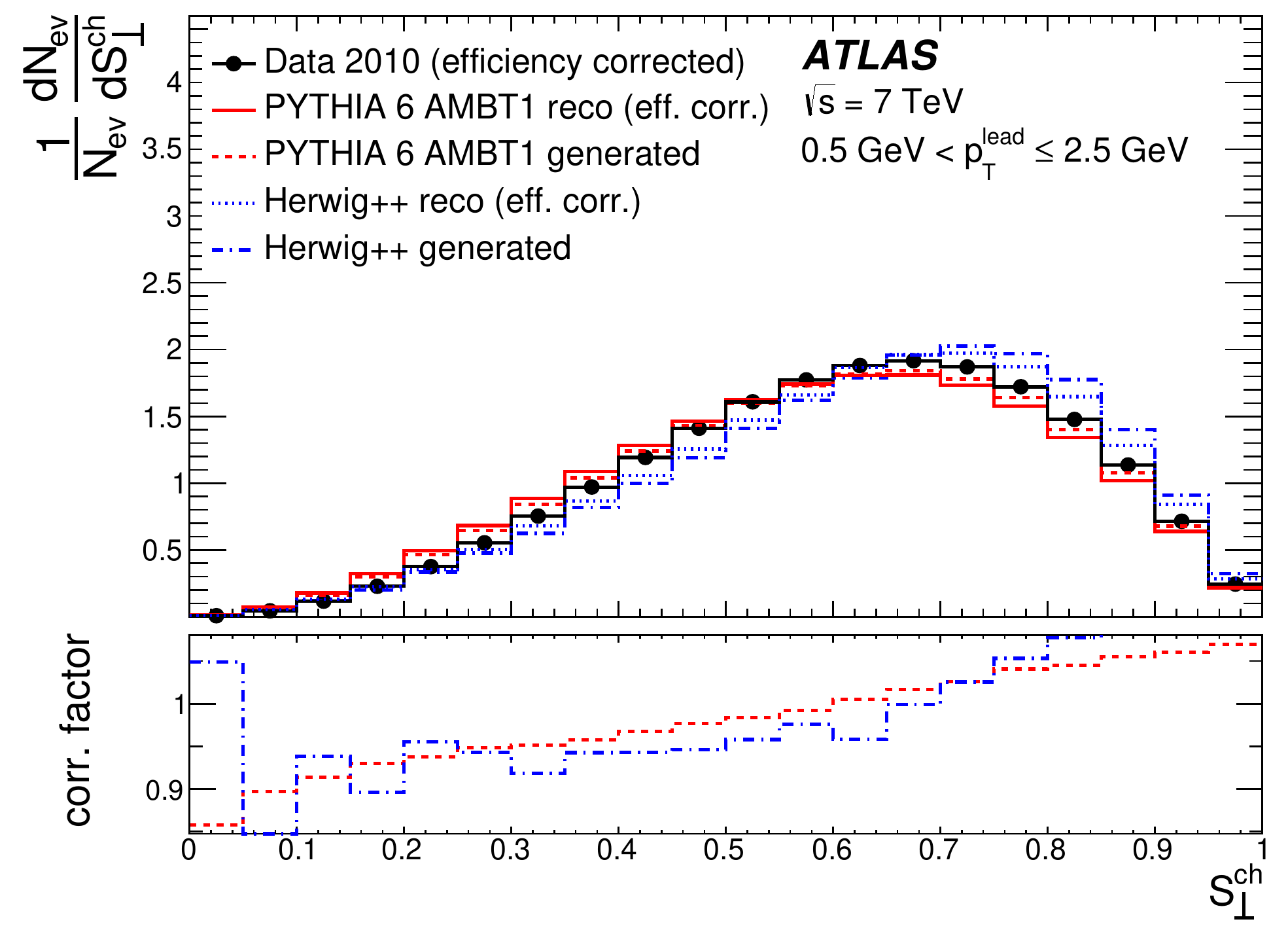}
	    \label{fig:U-sph}}

  
  \caption{The generated and reconstructed MC distributions of the complement of transverse thrust, the 
  thrust minor and the transverse sphericity are shown in the top part of each plot for the lowest
  \ptlead range.
         The correction factors are shown in the lower parts for \pythiasix AMBT1 and the \herwigpp default tune. 
          The data are shown with only the efficiency corrections and statistical uncertainties. Where not visible, the statistical error is smaller than the marker size.} 
  \label{fig:U3}
\end{figure}

The correction factors are calculated using the two different models implemented in 
\pythiasix AMBT1  and \herwigpp.
This correction accounts for bin-by-bin migrations and tracking inefficiencies.
For each distribution, the unfolding factor is typically within $\pm10\%$ of unity for most of the range.
It is very close to unity for the average values, except at the highest \ptsum .
The difference from unity becomes more pronounced at the statistically limited edges of the distributions.
The correction factors for the inclusive distributions of the three event shape observables
are shown in the bottom panels of~\FigRef{fig:U3}
for the two MC event generators mentioned above. 
Although the two MC generators have different distributions, the bin-by-bin correction factors are similar.

\section{Systematic uncertainties}
\label{sec:syst}

Systematic uncertainties on the measured distributions
are assessed with the following sources of uncertainty included:

\begin{description}
\item [Tracking] The largest of the systematic uncertainties for the tracking inefficiency~\cite{:2010ir}
 is found to be due to 
the material description in the inner detector.
This is determined to produce a relative uncertainty of $2\%$ in the efficiency in the barrel region, rising to $\sim$ $7\%$ for $2.3 < \etamod < 2.5$.
The contribution of the propagated
uncertainty is found to be less than $1\%$
of the content in each bin of the shape
distributions.

\item[Bin-by-bin correction model dependence] The remaining contributions to the overall systematic uncertainty result from
the specific correction method used in this analysis.  The bin-by-bin
corrections in general depend on the number of charged
particles and their \pt~ distributions, so there is some dependence on the event
generators. 
In order to estimate this uncertainty, it is necessary to compare different plausible
event generators, which deviate significantly from each other, but still give predictions  
close to the data.
The corrected results using the two very different 
\pythiasix AMBT1  and \herwigpp models are compared.
As these two generators use very different soft-QCD models the difference is assigned as 
a systematic uncertainty. 
The generated and reconstructed distributions are shown 
in ~\FigRef{fig:U3} for the two MC event generators and compared with the detector-level data.
%
\item[Statistical uncertainty of bin-by-bin correction] In addition to the model-dependent uncertainty in the
bin-by-bin correction, there is also a statistical uncertainty due to the finite
size of the MC sample. The statistical fluctuations of the \pythiasix AMBT1
correction factor are found to be negligible for most of each distribution,
increasing to a few percent in the tails of the distributions.
This is also added to the overall systematic uncertainty estimate.
\end{description}

The systematic uncertainty due to the small number
of residual multiple proton--proton interactions is estimated to be negligible.

All the above mentioned systematic uncertainties are added in quadrature.
\TabRef{tab:sysSummary} lists representative values
for the various contributions to the systematic
uncertainty in the content of each bin
for all the event shape observables
away from the edges of the distributions. 

\begin{table}[h]
  \singlespacing
  \caption{Summary of systematic uncertainties in $\%$.}
  \begin{center}
    \begin{tabular}{lrr}
      \toprule
      Trigger and vertex efficiency &  $< 0.1$  \\
      Track reconstruction & $0.1 \textendash 0.5$ \\
      Correction model difference  & $1\textendash5$  \\
      \pythia correction stat. uncertainty & $0.1\textendash2$ \\
      \midrule
      {\bf Total systematic uncertainty} &  $1 \textendash 5$     \\
      \bottomrule          
    \end{tabular}
    \label{tab:sysSummary}
  \end{center}
\end{table}

\section{Results and discussion}
\label{sec:results}

\begin{figure*}
  \begin{center}
    
      \subfigure[]{
        \includegraphics[width=0.95\columnwidth]{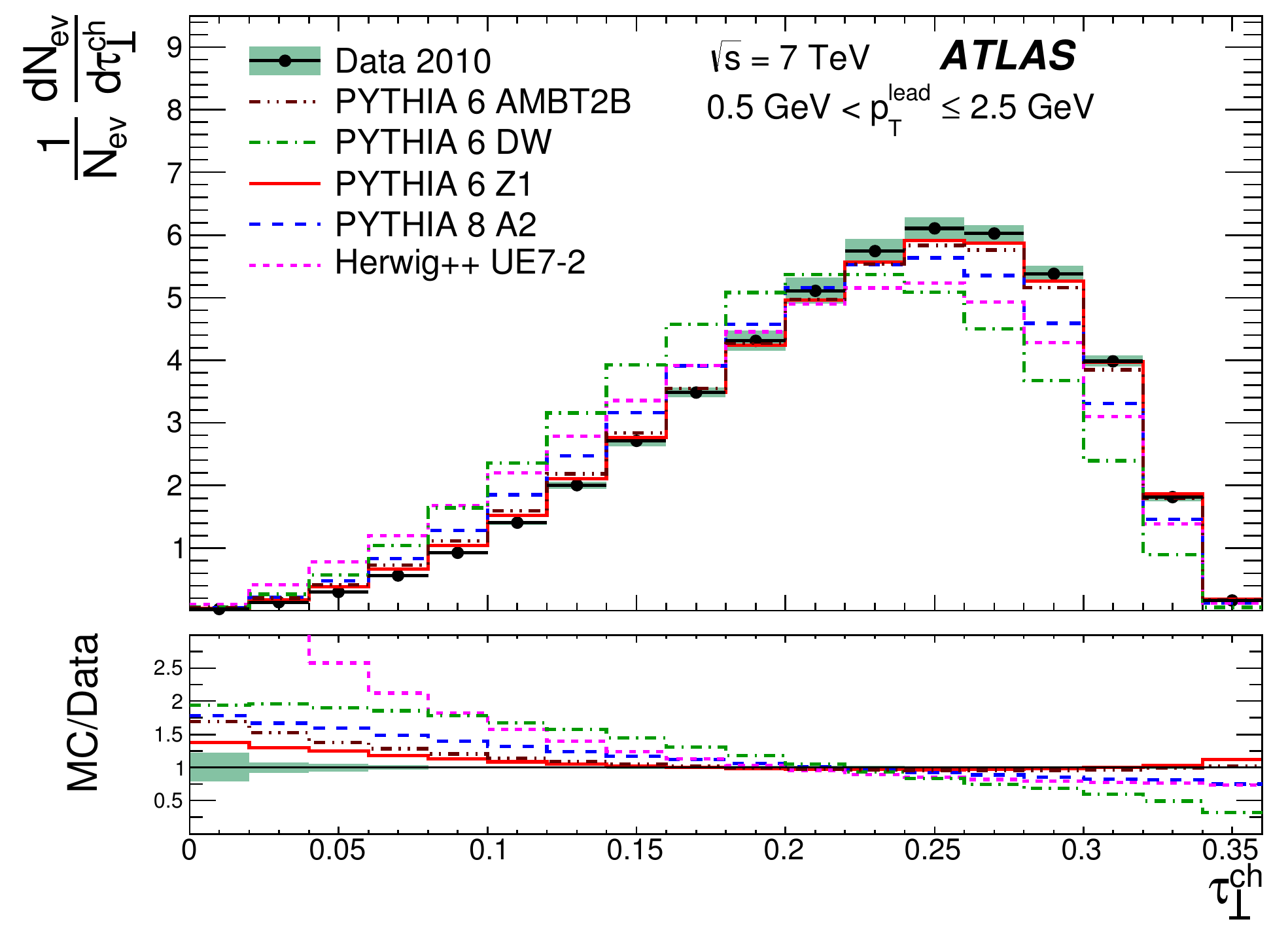}
        \label{fig:G-thr0}}
      \subfigure[]{
        \includegraphics[width=0.95\columnwidth]{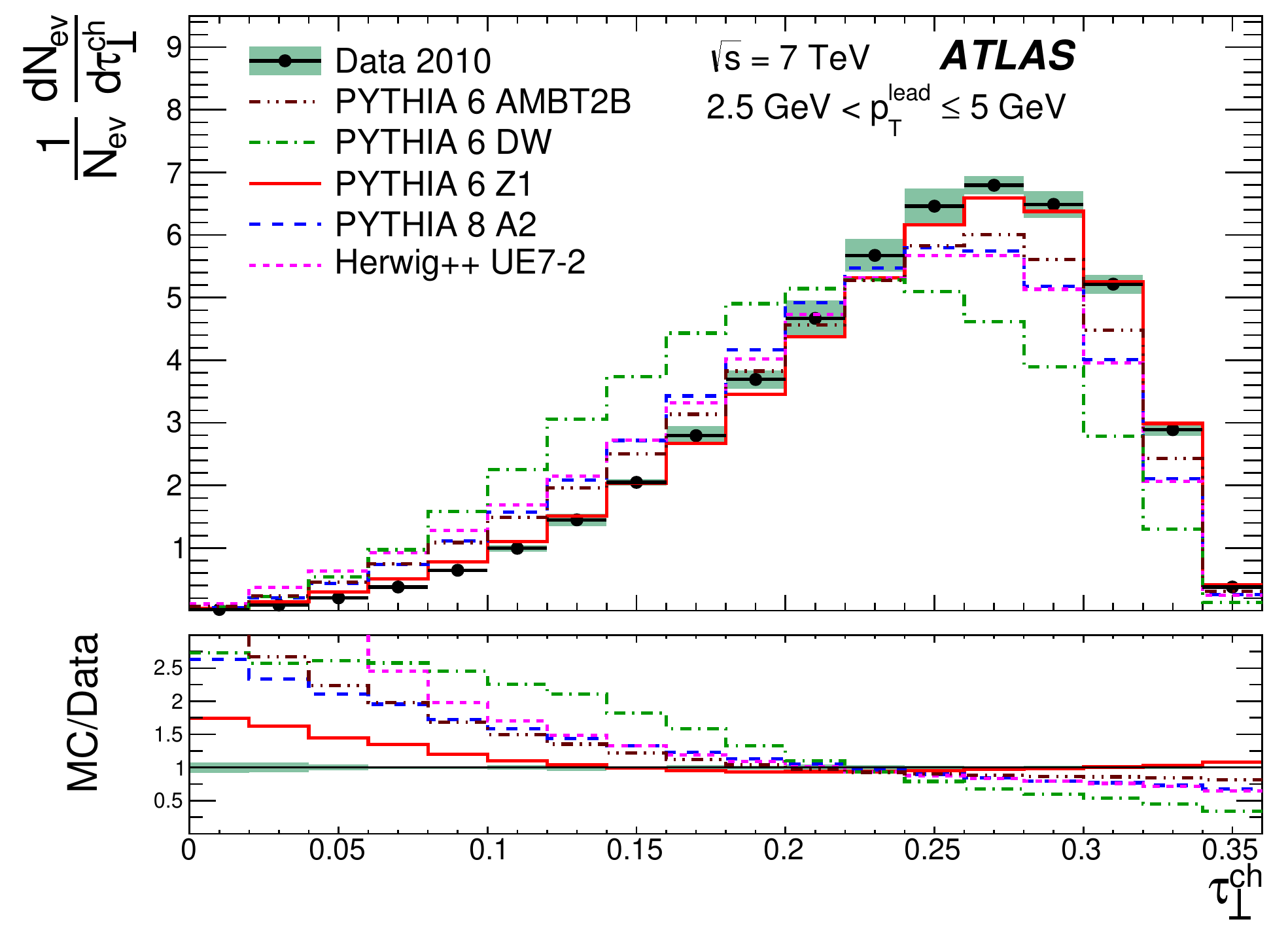}
        \label{fig:G-thr1}}
    
	  \subfigure[]{
	    \includegraphics[width=0.95\columnwidth]{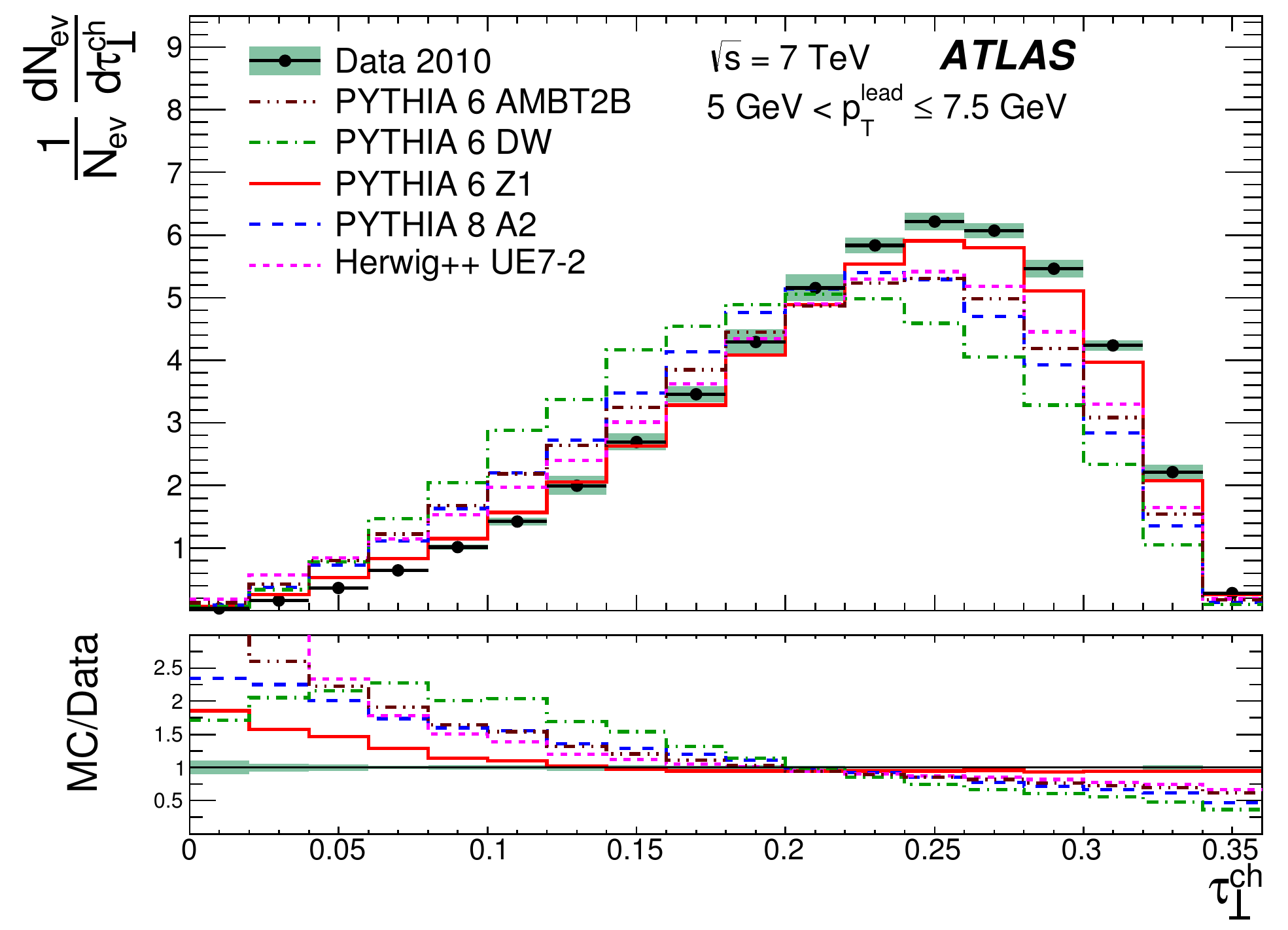}
	    \label{fig:G-thr2}}
	  \subfigure[]{
	    \includegraphics[width=0.95\columnwidth]{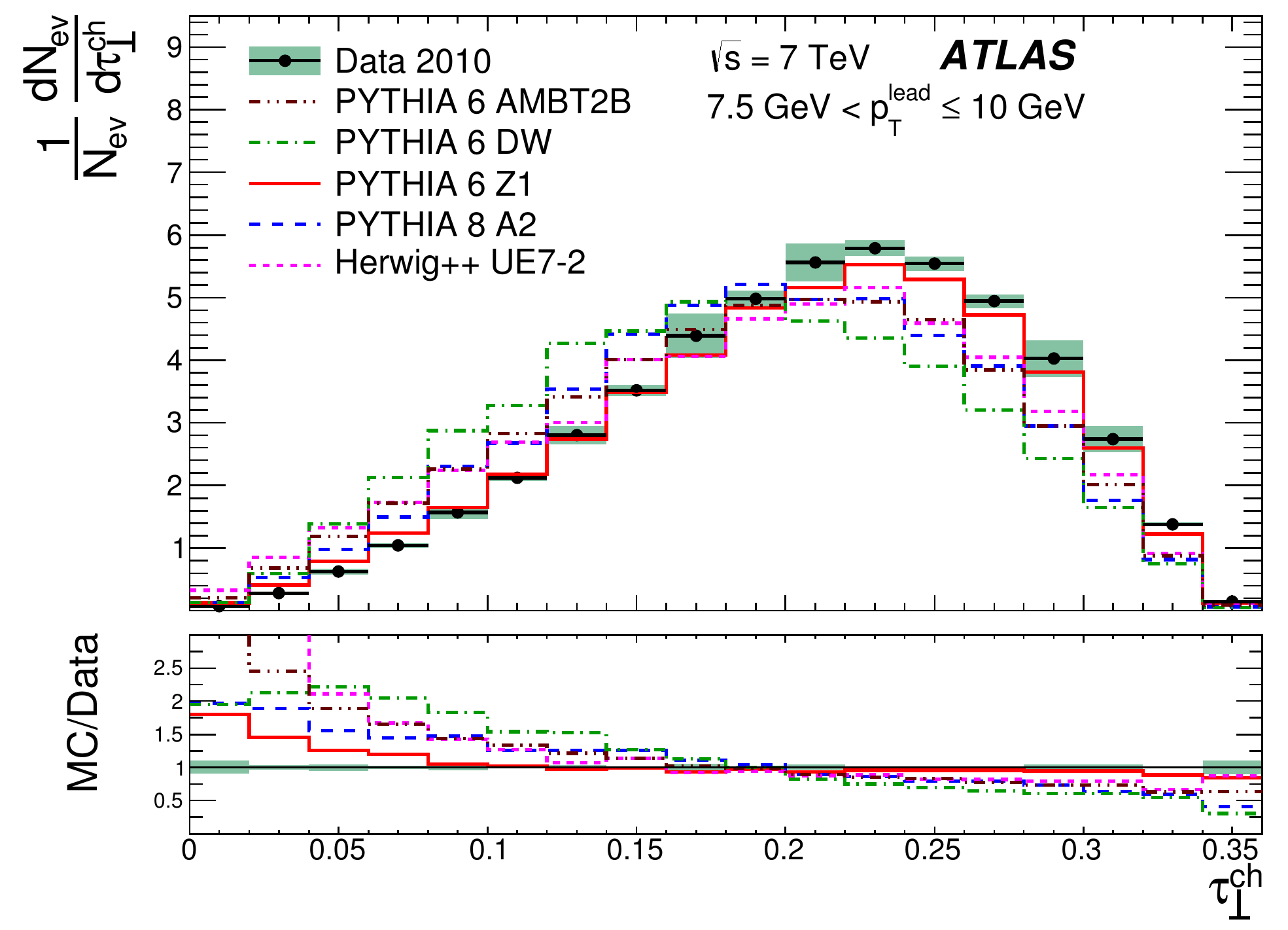}
	    \label{fig:G-thr3}}

	\centering
	  \subfigure[]{
	    \includegraphics[width=0.95\columnwidth]{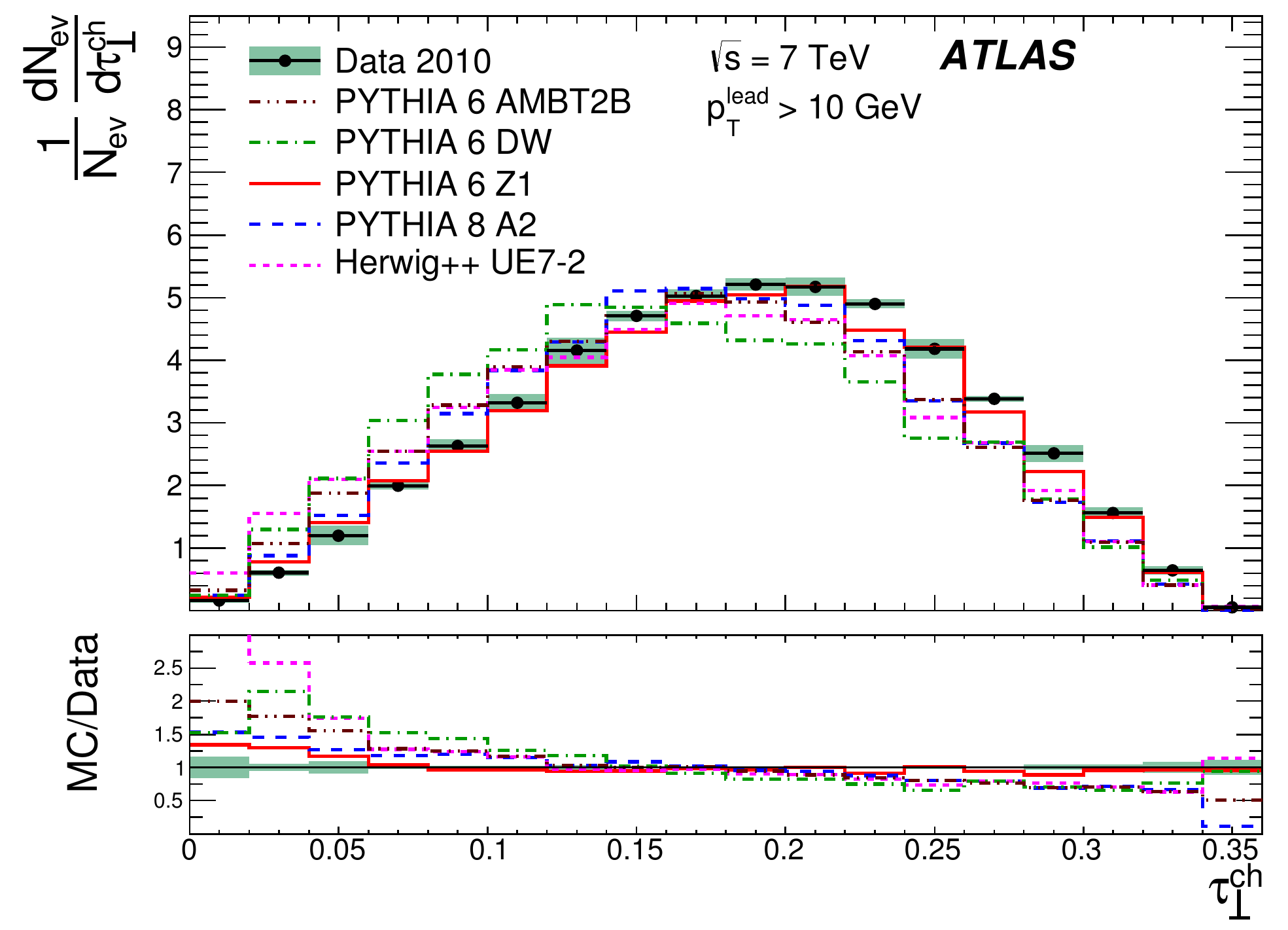}
	    \label{fig:G-thr4}}

  \end{center}
  
  \caption{Normalized distributions of the complement of transverse thrust using at least six
   charged particles with $\pT > 0.5$ GeV and $|\eta| < 2.5$ for 
  different requirements on the transverse momentum of the leading charged particle, \ptlead. 
  The error bars show the statistical uncertainty while the shaded area shows the combined statistical and systematic uncertainty. Where not visible, the statistical error is smaller than the marker size.  }
 
  \label{fig:G-thr}
\end{figure*}

\begin{figure*}
  \begin{center}
    
      \subfigure[]{
        \includegraphics[width=0.95\columnwidth]{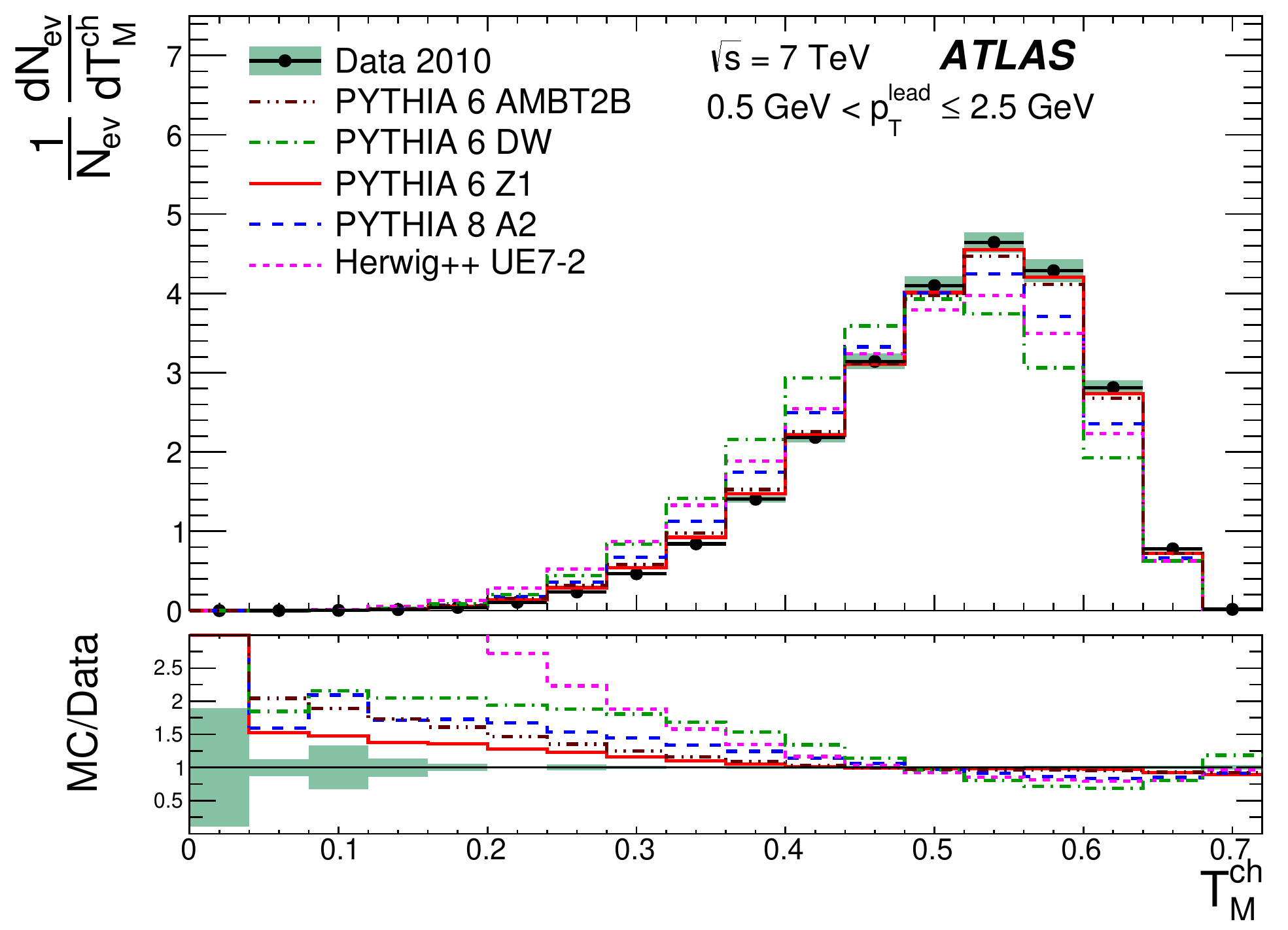}
        \label{fig:G-tm0}}
      \subfigure[]{
        \includegraphics[width=0.95\columnwidth]{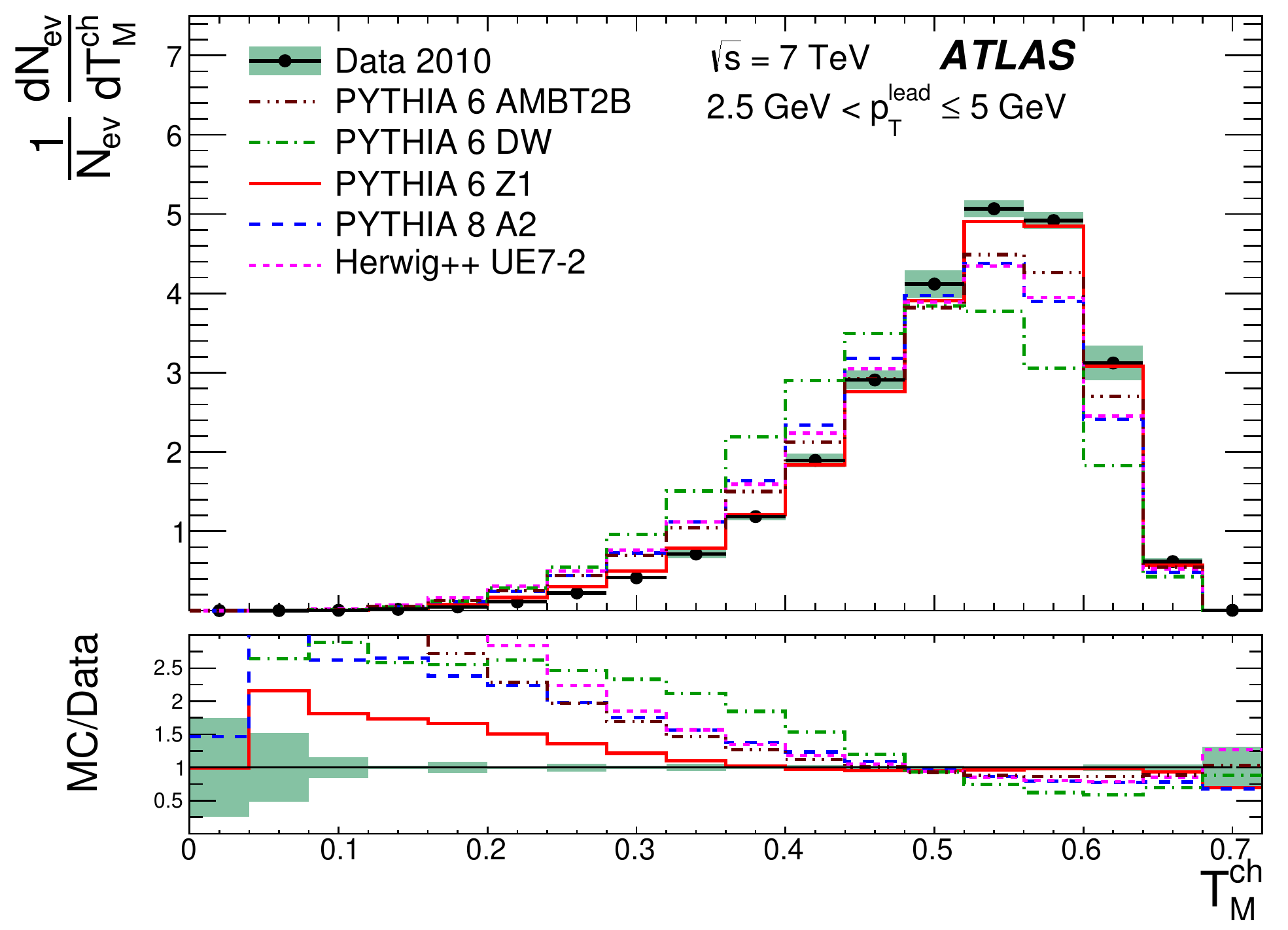}
        \label{fig:G-tm1}}
    
	  \subfigure[]{
	    \includegraphics[width=0.95\columnwidth]{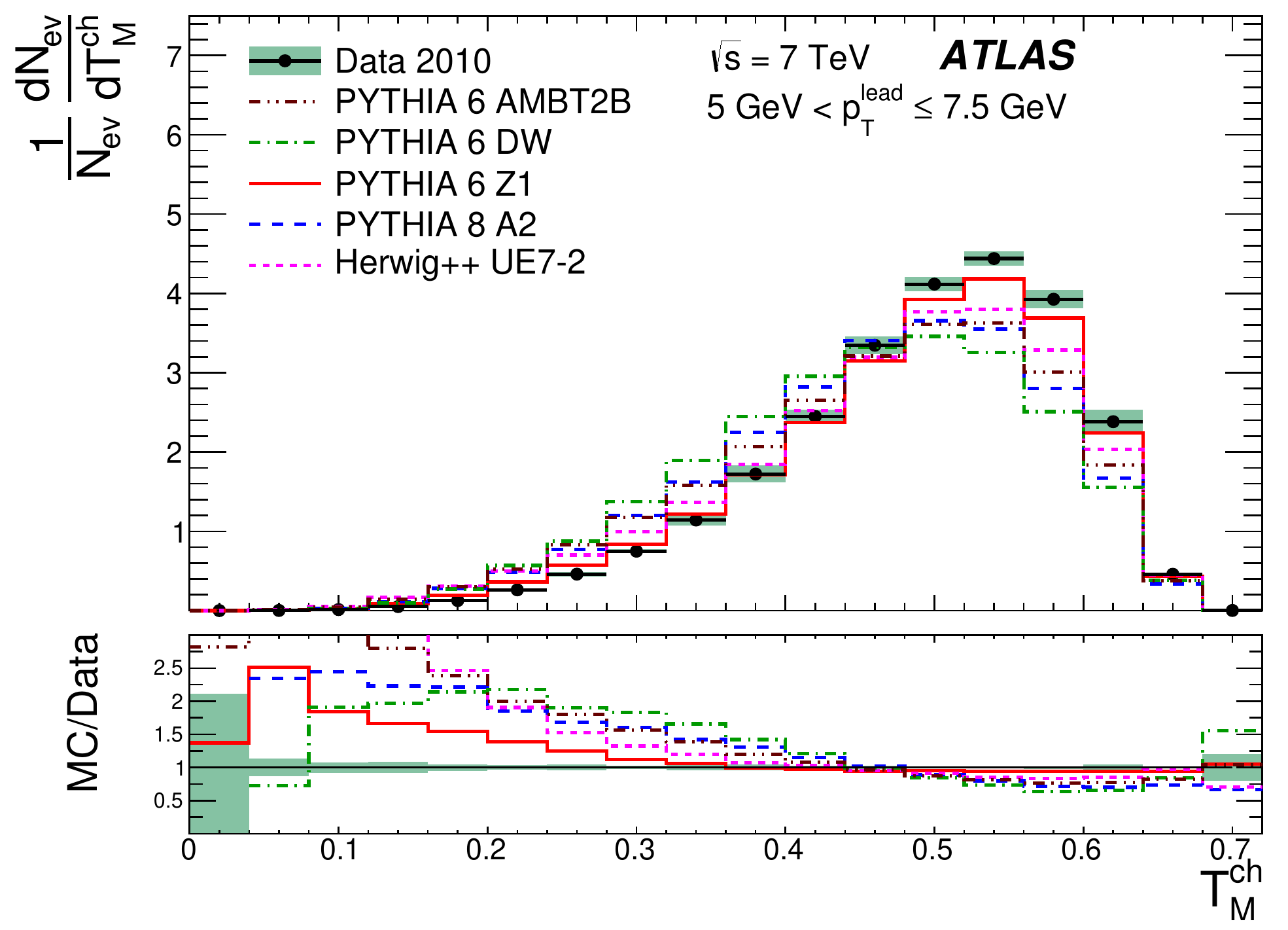}
	    \label{fig:G-tm2}}
	  \subfigure[]{
	    \includegraphics[width=0.95\columnwidth]{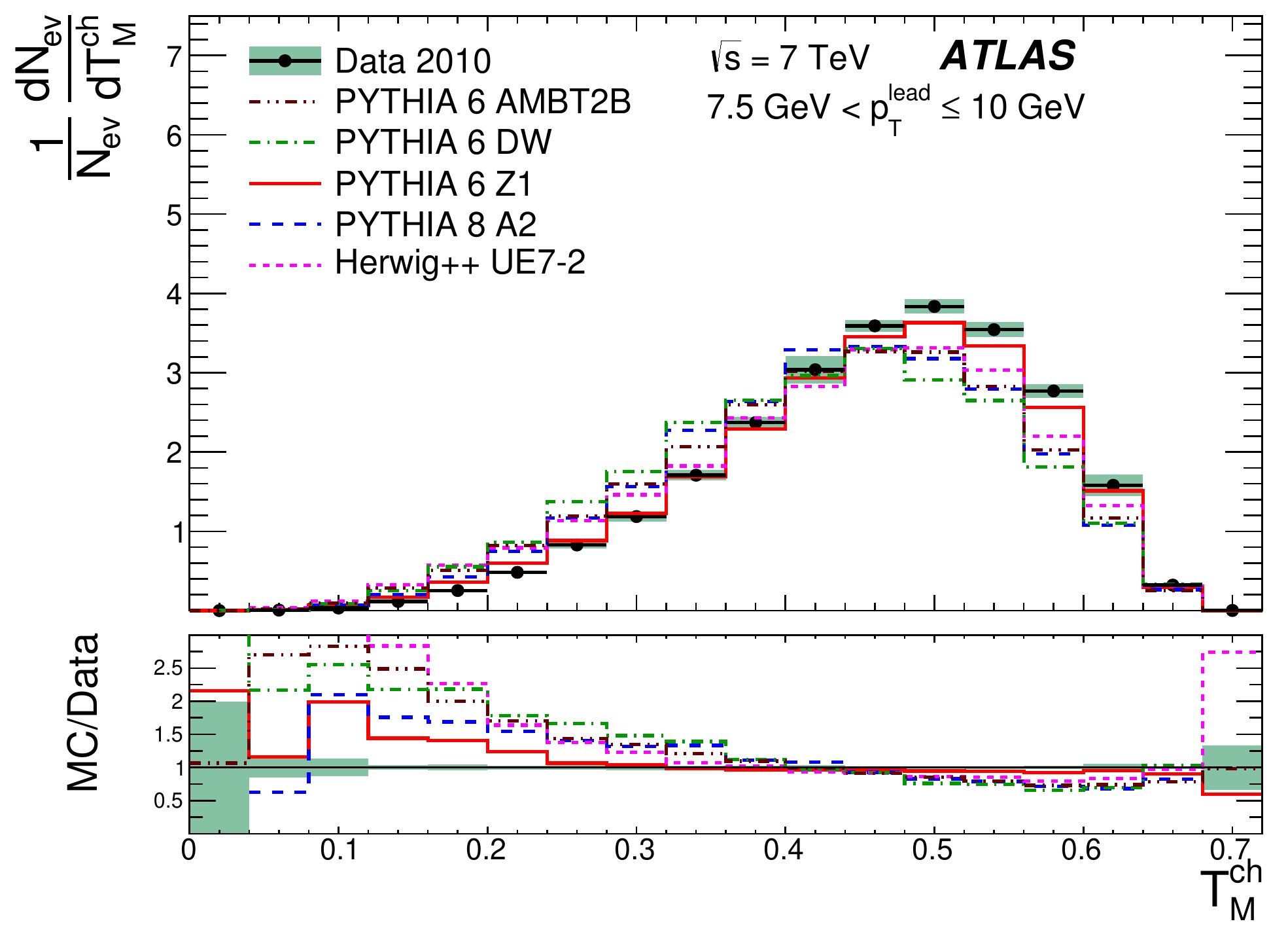}
	    \label{fig:G-tm3}}

	  \subfigure[]{
	    \includegraphics[width=0.95\columnwidth]{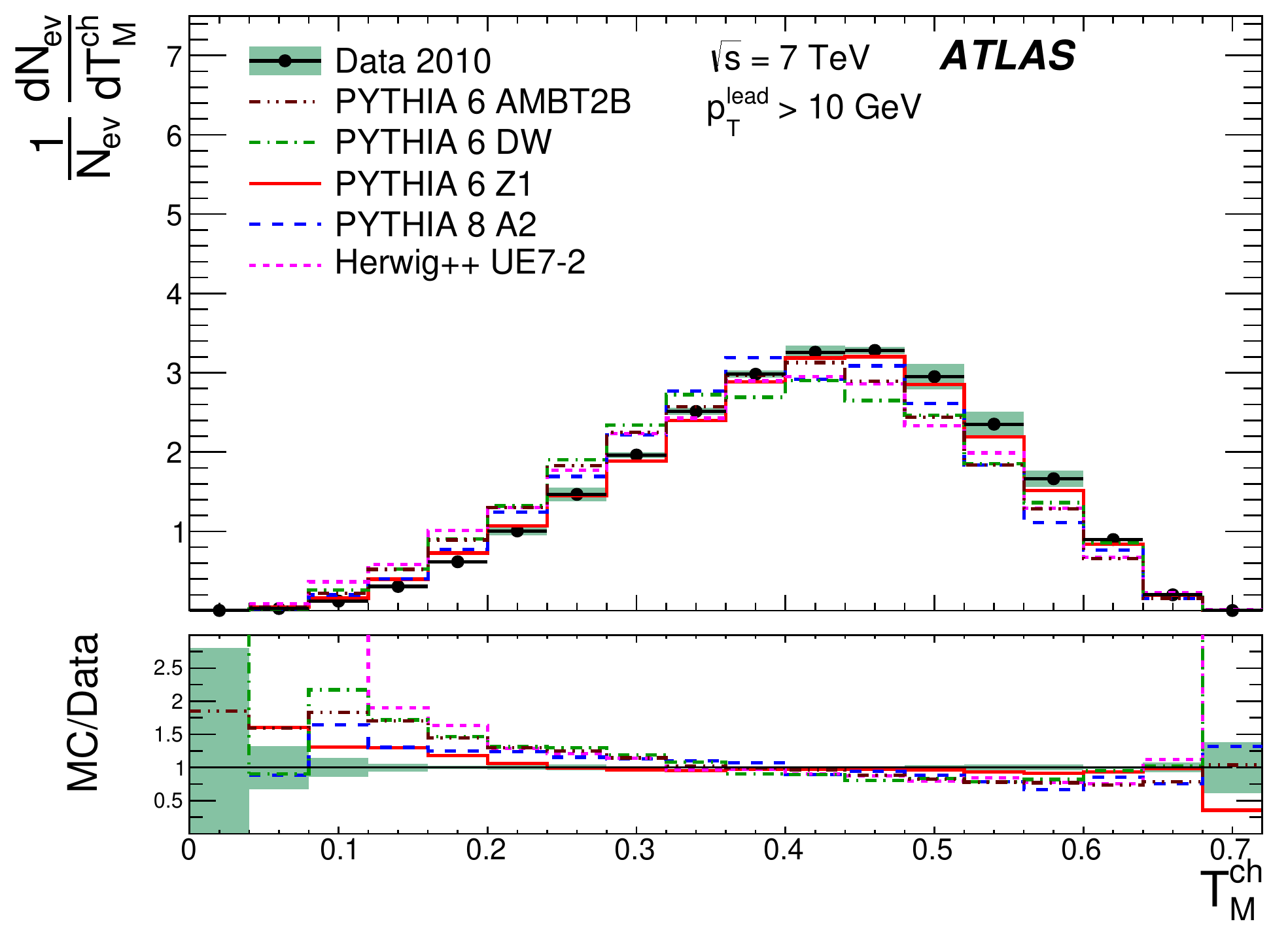}
	    \label{fig:G-tm4}}

  \end{center}
  
  \caption{Normalized distributions of  transverse thrust minor using at least six
   charged particles with $\pT > 0.5$ GeV and $|\eta| < 2.5$ for 
  different requirements on the transverse momentum of the leading charged particle, \ptlead . 
  The error bars show the statistical uncertainty while the shaded area shows the combined statistical and systematic uncertainty. Where not visible, the statistical error is smaller than the marker size.}
 
  \label{fig:G-tm}
\end{figure*}

\begin{figure*}
  \begin{center}
    
      \subfigure[]{
        \includegraphics[width=0.95\columnwidth]{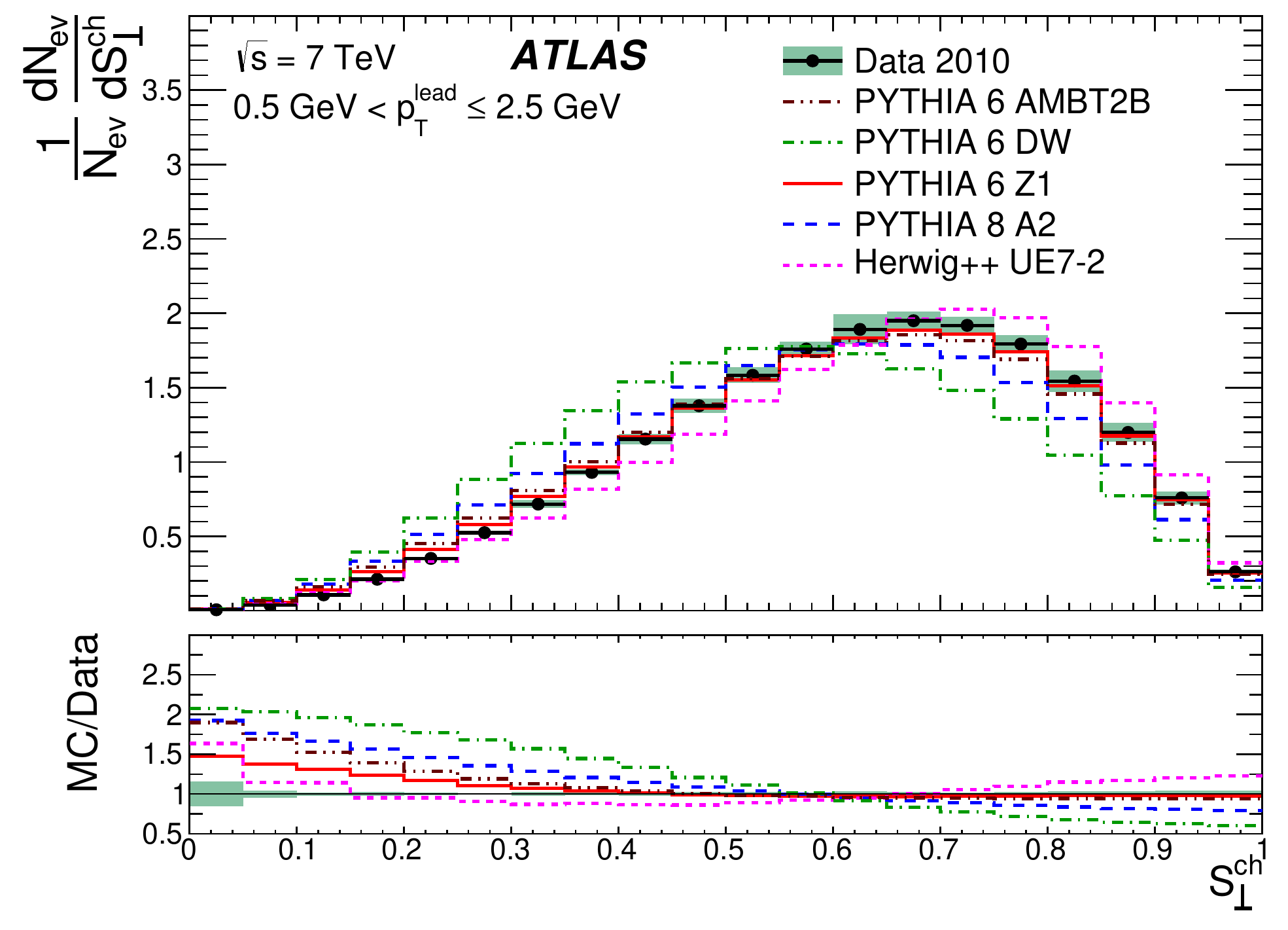}
        \label{fig:G-sph0}}
      \subfigure[]{
        \includegraphics[width=0.95\columnwidth]{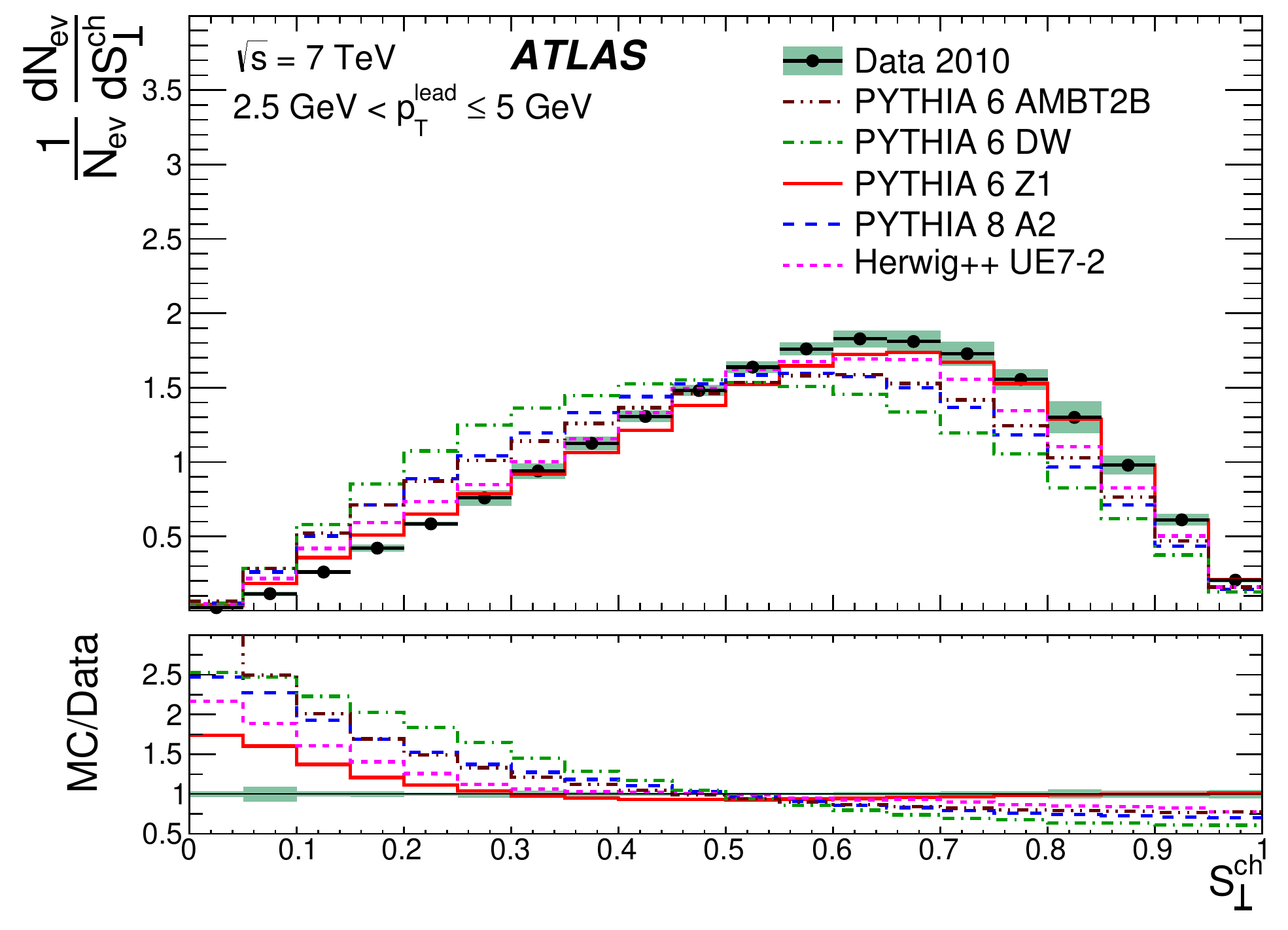}
        \label{fig:G-sph1}}
    
	  \subfigure[]{
	    \includegraphics[width=0.95\columnwidth]{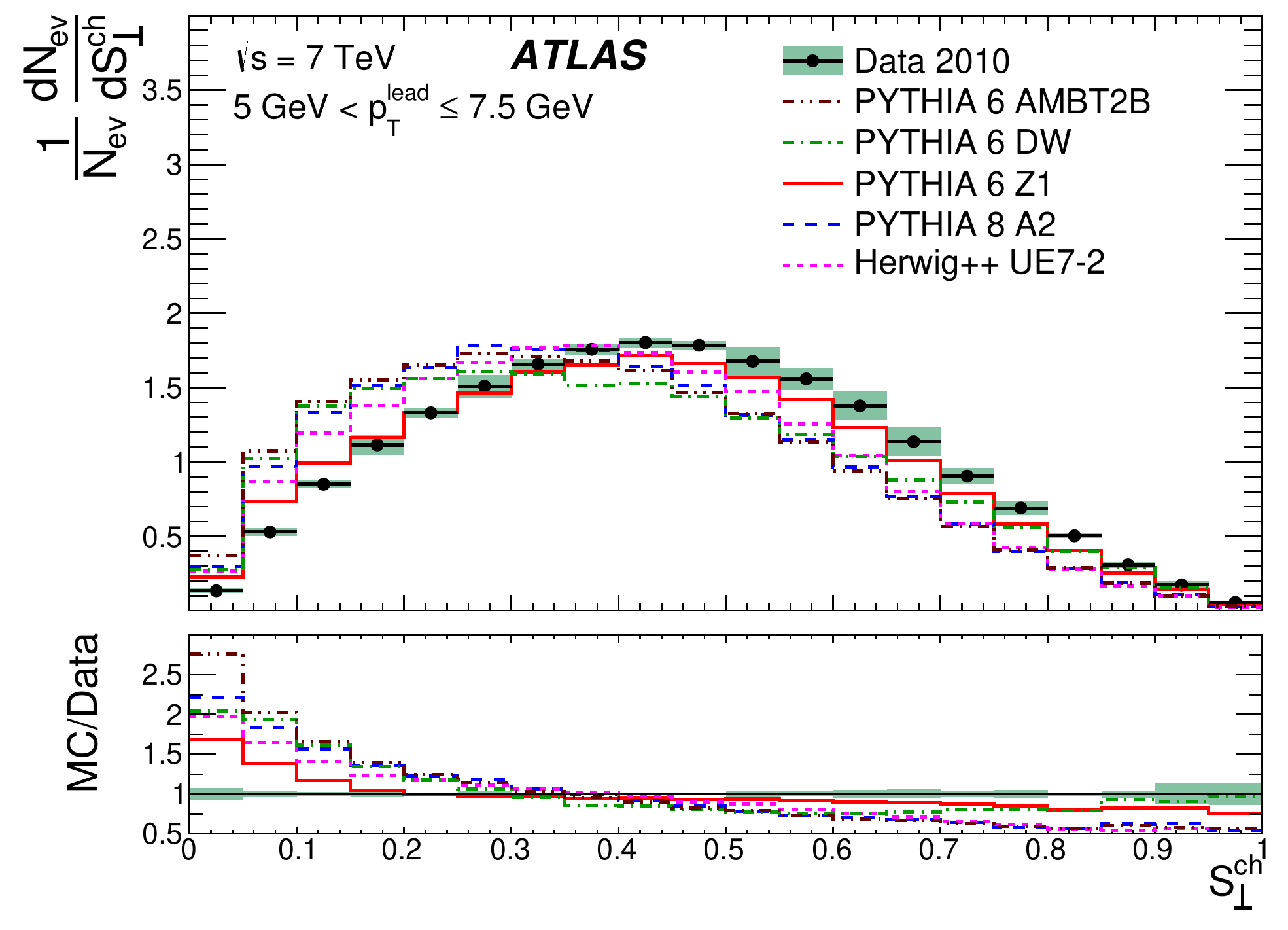}
	    \label{fig:G-sph2}}
	  \subfigure[]{
	    \includegraphics[width=0.95\columnwidth]{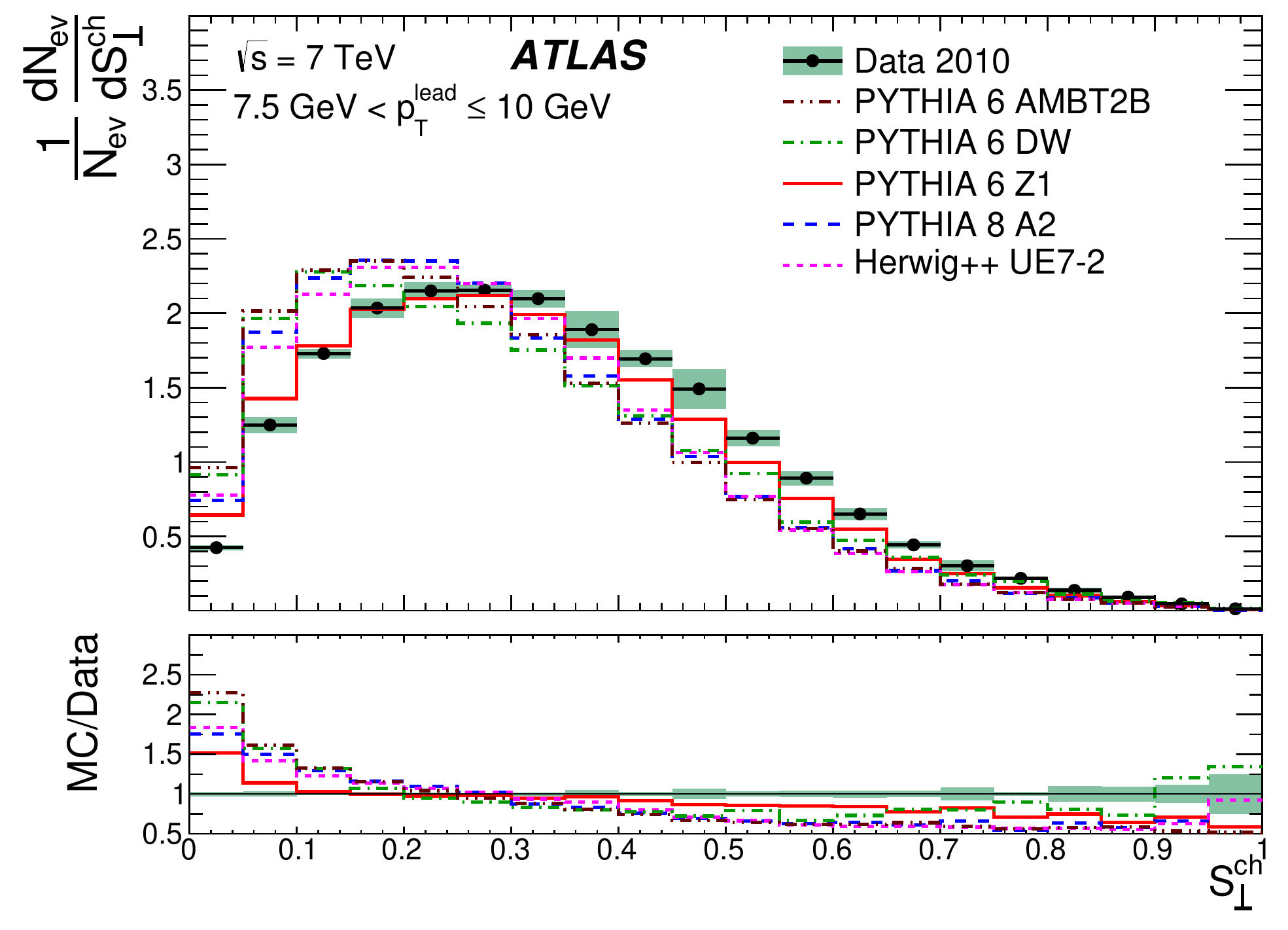}
	    \label{fig:G-sph3}}

	  \subfigure[]{
	    \includegraphics[width=0.95\columnwidth]{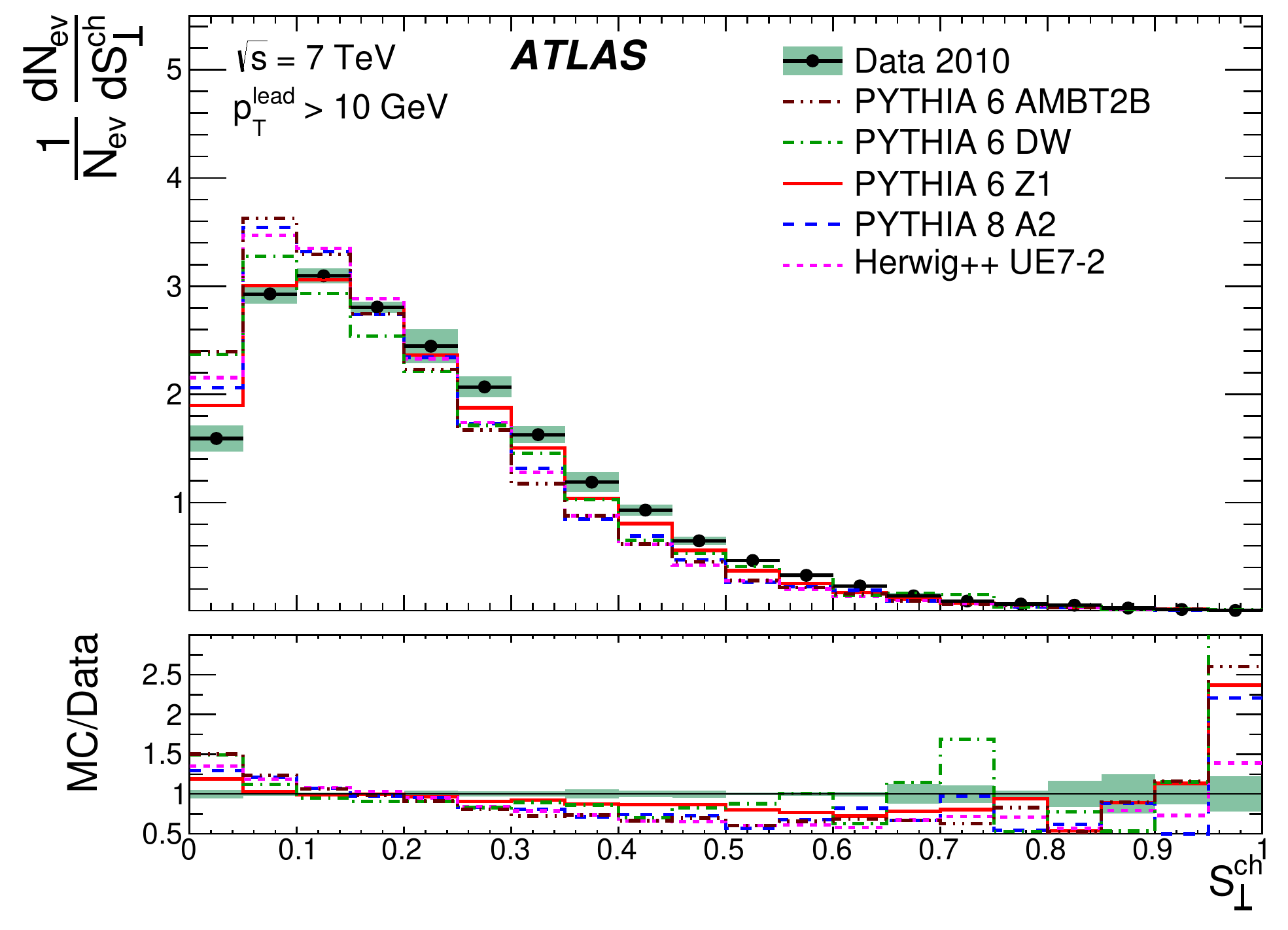}
	    \label{fig:G-sph4}}

  \end{center}
  
  \caption{Normalized distributions of  transverse sphericity  using at least six
   charged particles with $\pT > 0.5$ GeV and $|\eta| < 2.5$ for 
  different requirements on the transverse momentum of the leading charged particle, \ptlead . 
  The error bars show the statistical uncertainty while the shaded area shows the combined statistical and systematic uncertainty. Where not visible, the statistical error is smaller than the marker size.
  }
 
  \label{fig:G-sph}
\end{figure*}

\begin{figure*}
  \begin{center}
    
      \subfigure[]{
        \includegraphics[width=0.95\columnwidth]{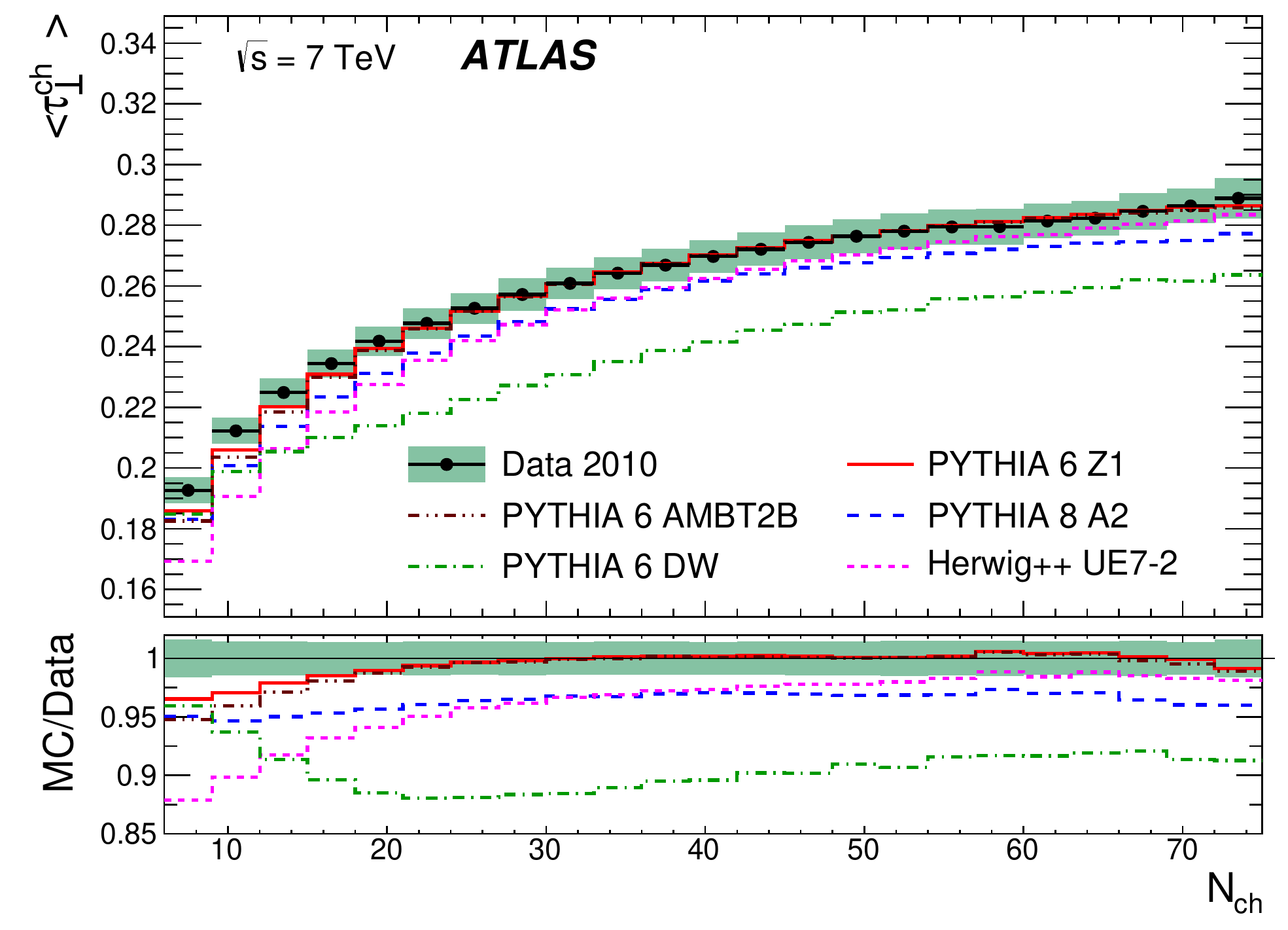}
        \label{fig:G-thrnch}}
      \subfigure[]{
        \includegraphics[width=0.95\columnwidth]{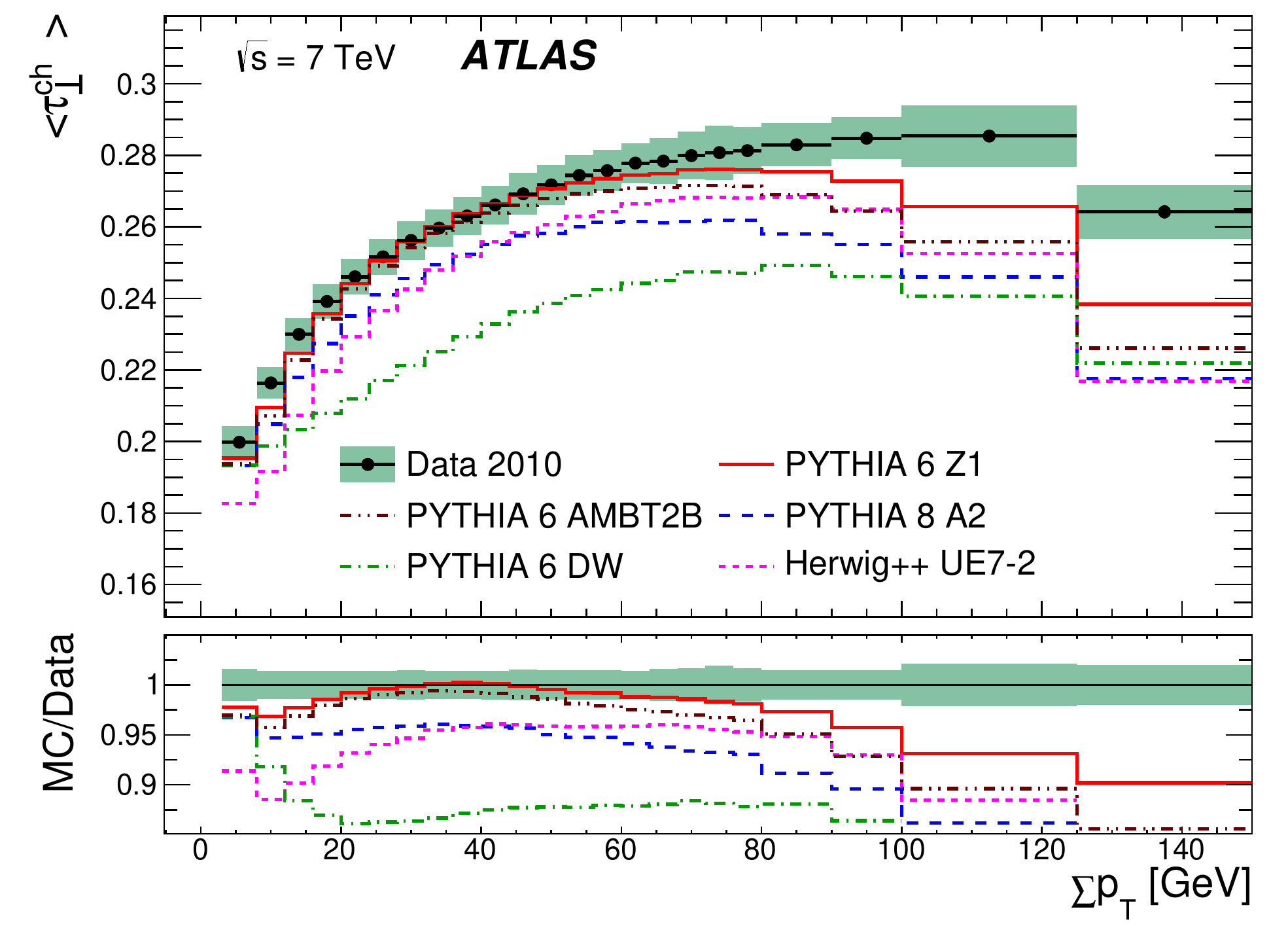}
        \label{fig:G-thrptsum}}
    
	  \subfigure[]{
	    \includegraphics[width=0.95\columnwidth]{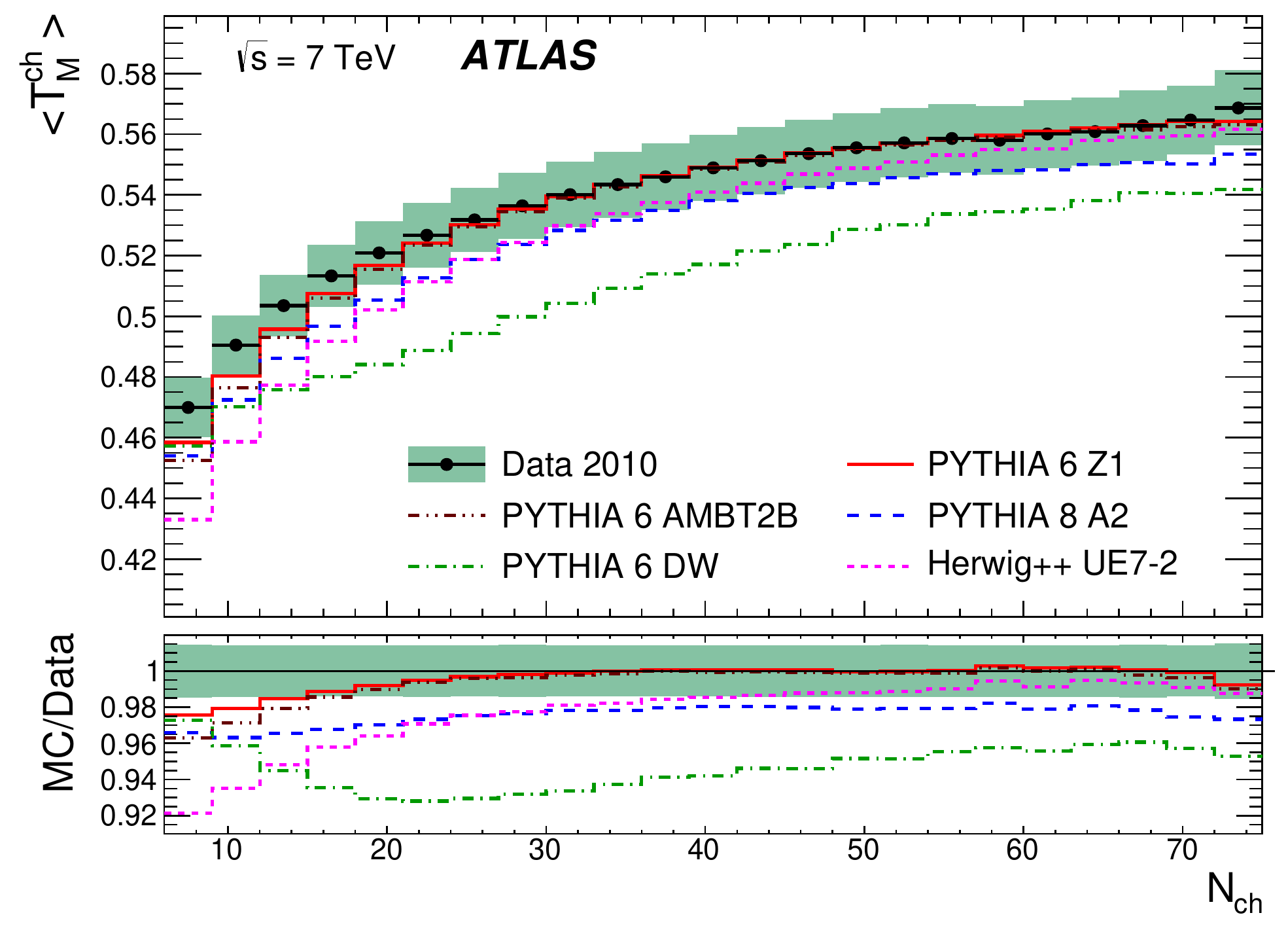}
	    \label{fig:G-tmnch}}
	  \subfigure[]{
	    \includegraphics[width=0.95\columnwidth]{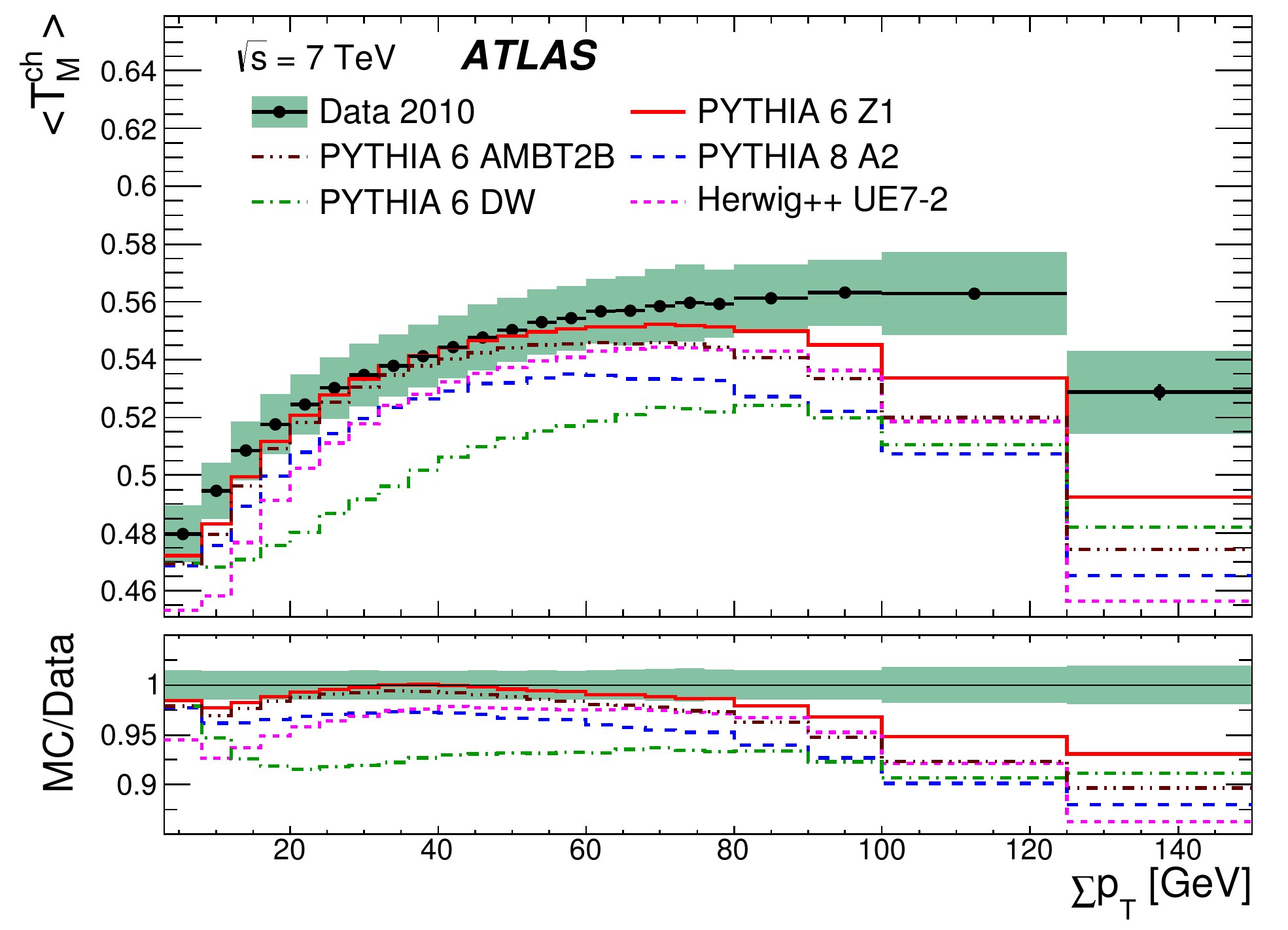}
	    \label{fig:G-tmptsum}}

	  \subfigure[]{
	    \includegraphics[width=0.95\columnwidth]{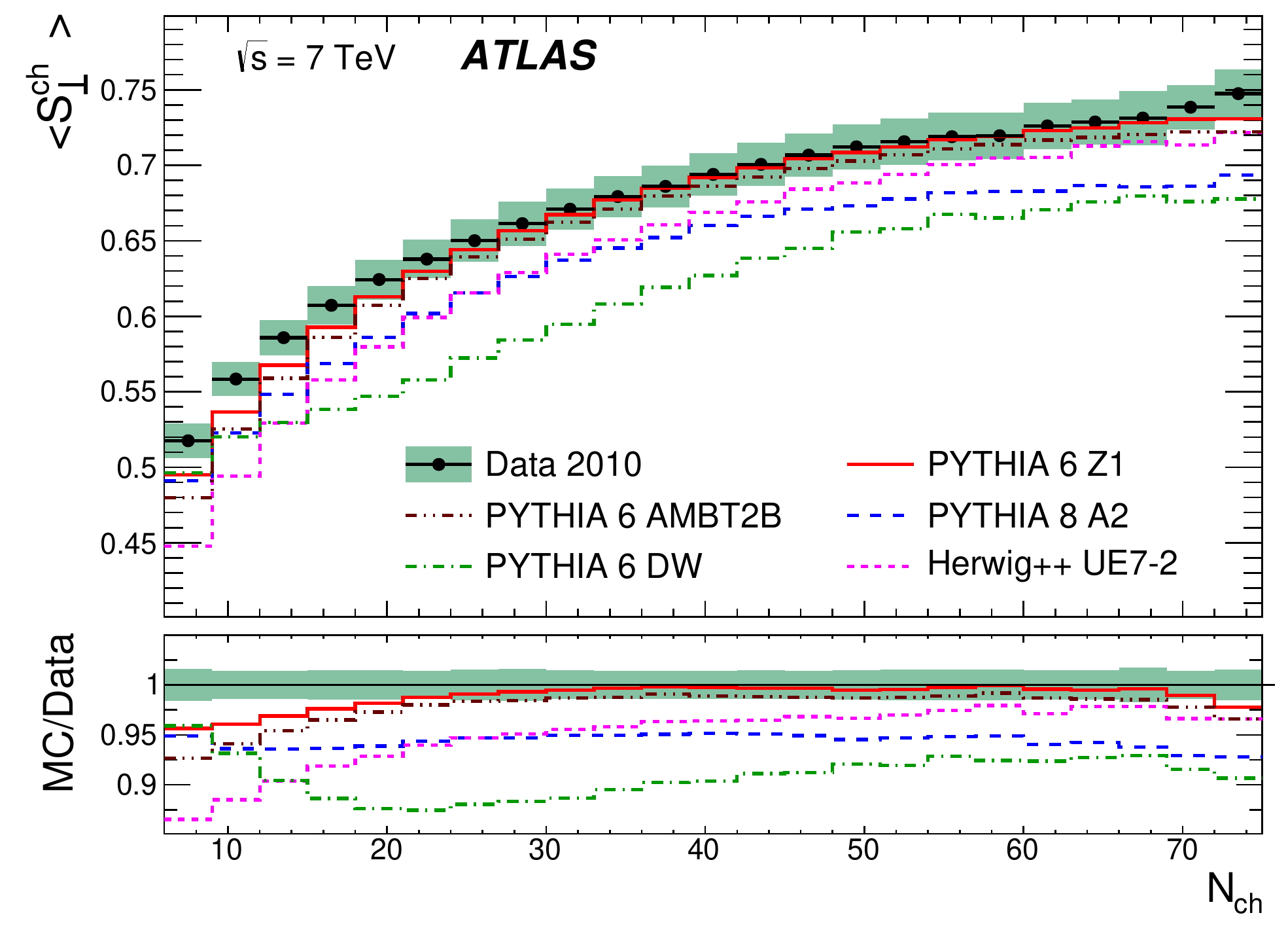}
	    \label{fig:G-sphnch}}
	  \subfigure[]{
	    \includegraphics[width=0.95\columnwidth]{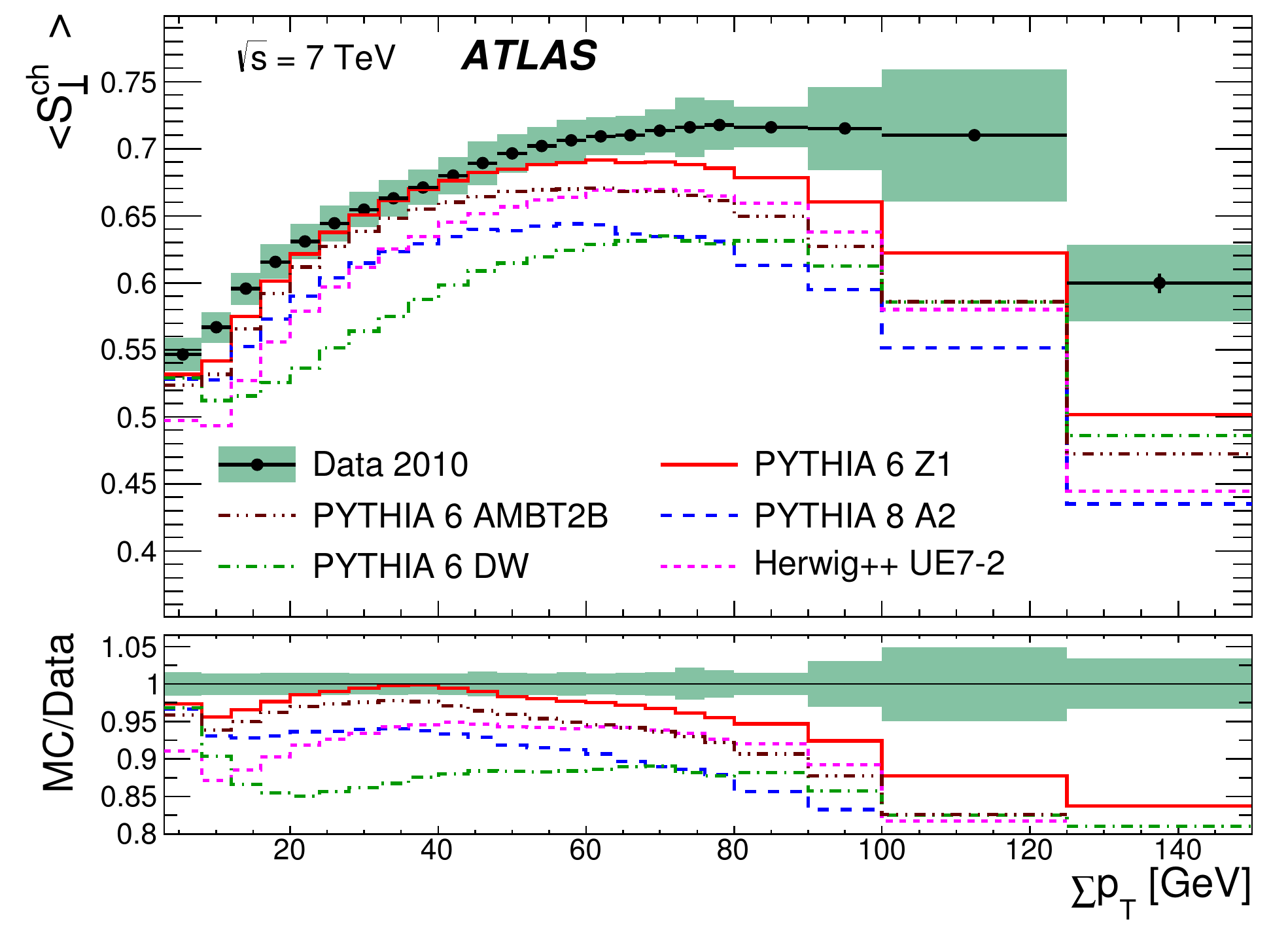}
	    \label{fig:G-sphptsum}}

  \end{center}
  
  \caption{Mean values of the complement of transverse thrust, transverse
   thrust minor and transverse sphericity (top to bottom) using at least six charged particles 
   with $\pT > 0.5$ GeV and $|\eta| < 2.5$ versus charged particle 
   multiplicity of the event (left) and versus 
   charged particle transverse momentum scalar sum of the event (right).
   The error bars show the statistical uncertainty while the shaded area shows the combined statistical and systematic uncertainty. Where not visible, the statistical error is smaller than the marker size.
  }
  \label{fig:G-pro}
\end{figure*}

The distributions of the complement of the transverse thrust, thrust minor and transverse sphericity are 
presented in Figs.~\ref{fig:G-thr}--\ref{fig:G-sph},
in different \ptlead ranges.
The behavior of the average values of the shape variables as functions of 
the charged particle multiplicity, \Nchg, and transverse momentum scalar 
sum, \ptsum, is presented in Fig.~\ref{fig:G-pro}. Predictions from the
\pythiasix AMBT2B,
\pythiasix DW, \pythiasix Z1, \pythiaeight A2
and \herwigpp UE7-2 models are also shown.
AMBT2B is chosen instead of AMBT1, which was
used to correct the data  back to the particle level because it shows a slight improvement
in reproducing the
distributions of charged particle transverse momentum and multiplicity~\cite{:mc11}.

The distributions shown in Figs.~\ref{fig:G-thr} -- \ref{fig:G-sph} indicate a prevalence of 
spherical events in the lower \ptlead ranges. A slight shift toward less
spherical events and a broadening of the distributions is observed for events
starting with \mbox{\ptlead $>7.5$~GeV} in Fig.~\ref{fig:G-thr3} for $\tau_{\perp}^{\text{ch}}$ and 
in Fig.~\ref{fig:G-tm3} for $T_{\text{M}}^{\text{ch}}$. 
For both variables, a transition to less spherical events is seen
for \ptlead $>10$~GeV in Fig.~\ref{fig:G-thr4} and in Fig.~\ref{fig:G-tm4}.
The distribution of transverse sphericity is 
more sensitive to the increase of \ptlead, and shows a marked shift toward less spherical
events starting at \ptlead $> 5.0$~GeV in Fig.~\ref{fig:G-sph2}.
The average value of the distributions, the RMS width and the skewness of the distributions
are given in \TabRef{tab:results}, which supports this observation.
Mean values of the complement of transverse thrust and the transverse 
thrust minor are observed to initially rise with increasing
\ptlead, with their maximum value in the range $2.5 <
\ptlead< 5 \GeV$, before decreasing.
%
A similar trend is observed by the ALICE Collaboration, which has measured the transverse sphericity distribution
selecting charged particles with $|\eta| < 0.8$, in inelastic $\unit{7}{\TeV}$ $pp$ collisions~\cite{Abelev:2012sk}.

Overall, the \pythiasix tune Z1, tuned to the underlying event distributions
at the LHC, agrees the best with most of the distributions.
The \pythiasix DW tune predictions are consistently furthest from the data,
as seen in the $\tau_{\perp}^{\text{ch}}$ and $T_{\text{M}}^{\text{ch}}$ distributions.
This is not unexpected as DW is tuned to reproduce the Tevatron data and
does not agree with the charged particle multiplicity and \pT\
distributions in LHC data~\cite{:2010ir}. 
However it performs similarly to other 
models/tunes for the $S_{\perp}^{\text{ch}}$
distribution
in intermediate to high \ptlead values,
as is seen in Fig.~\ref{fig:G-sph2}--Fig.~\ref{fig:G-sph4}.
The AMBT2B tune, which is based on minimum-bias LHC data, shows better agreement
for the lowest \ptlead distributions than for the intermediate \ptlead distributions,
as is seen in Fig.~\ref{fig:G-thr0} and in Fig.~\ref{fig:G-tm0}.
Compared to the \pythiasix AMBT2B tune, the predictions of the \pythiaeight A2 and \herwigpp UE7-2 tunes show better agreement with the data in the intermediate to high \ptlead ranges.
The UE7-2 tune, based like Z1 on LHC underlying event data, 
is expected to perform better in events characterized by 
a hard scatter, resulting in higher \ptlead values. 
However, the minimum-bias A2 tune shows
a similar or slightly better level of agreement with data for the high \ptlead distributions,
possibly indicating that the improved MPI modeling compared to \pythiasix tunes
does play a role.
All models tend to better reproduce the data selected with the higher \ptlead ranges.

The mean values of event shape observables as functions of \Nchg and \ptsum are shown in \FigRef{fig:G-pro}.
They are seen to increase with \Nchg, but the increase is less marked at values of \Nchg above about 30.
For low values of \Nchg, the mean values of the event shape variables correspond to 
less spherical events, while the
average values for large multiplicity   
is largely consistent with the positions of the maxima of the corresponding distributions for the lowest \ptlead range.
A similar trend is seen for distributions as a function of \ptsum; however, for \ptsum over $100$ GeV, 
the mean starts to decrease again, indicative of a dijet topology. 
In general, the MC models predict fewer high-sphericity events than are seen in the data.
With the exception of \pythiasix DW, the MC models seem to predict 
the behavior with multiplicity   
reasonably well in \FigRef{fig:G-pro}.
However, the MC predictions are seen to differ in shape at very high \ptsum, where the decrease of mean values
happens in the MC predictions before the data.
The behavior of mean transverse sphericity as a function of multiplicity measured by the ALICE Collaboration~\cite{Abelev:2012sk}
exhibits a similar behavior to that observed here, with the data lying at values higher than predicted by the MC models.


\begin{table*}
 \caption{Mean, RMS and skewness for each event shape distribution is shown,
 in different intervals of \ptlead .
 Combined statistical and systematic uncertainty is shown, where the systematic
 uncertainty is obtained from the difference of unfolded results using \textsc{Pythia 6} and \textsc{Herwig\smaller++}
 MC predictions. 
 }
\begin{center}
\begin{tabular}[c]{rcl@{\hskip 0.75in}cc@{\hskip 0.5in}cc@{\hskip 0.5in}cc}
\toprule
&&&\multicolumn{6}{c}{\textbf{1 - Transverse Thrust}} \\
\multicolumn{3}{c}{\textbf{\ptlead range}}  & \multicolumn{2}{l}{\hspace{.25in}Mean} & \multicolumn{2}{l}{\hspace{.25in}RMS} & \multicolumn{2}{c}{Skewness} \\
\midrule
$ \unit{0.5}{\GeV} < $ & \ptlead  &$\leq \unit{2.5}{\GeV}$  & 0.227 & $\pm$ 0.002 & 0.064 & $\pm$ 0.008 & \textendash0.54 & $\pm$ 0.03 \\ 
$ \unit{2.5}{\GeV} < $ &\ptlead &  $\leq \unit{5.0}{\GeV}$& 0.240 & $\pm$ 0.006 & 0.062 & $\pm$ 0.001  & \textendash0.68 & $\pm$ 0.04 \\ 
$ \unit{5.0}{\GeV} < $ & \ptlead  & $\leq \unit{7.5}{\GeV}$ & 0.227 & $\pm$ 0.007 & 0.065 & $\pm$ 0.003 & \textendash0.55 & $\pm$ 0.04 \\ 
$ \unit{7.5}{\GeV} < $ & \ptlead & $\leq \unit{10}{\GeV}$ &  0.210 & $\pm$ 0.010 & 0.068 & $\pm$ 0.005 & \textendash0.36 & $\pm$ 0.09 \\
& \ptlead   & $ > \unit{10}{\GeV}$  &  0.185 & $\pm$ 0.011 & 0.070 & $\pm$ 0.006 & \textendash0.11 & $\pm$ 0.28 \\ 
\midrule
&&& \multicolumn{6}{c}{\textbf{Thrust Minor}} \\
\multicolumn{3}{c}{\textbf{\ptlead range}} & \multicolumn{2}{l}{\hspace{.25in}Mean} & \multicolumn{2}{l}{\hspace{.25in}RMS} & \multicolumn{2}{c}{Skewness} \\		
\midrule
$ \unit{0.5}{\GeV} < $ & \ptlead  &$ \leq \unit{2.5}{\GeV}$  &  0.508 & $\pm$ 0.002 & 0.090 & $\pm$ 0.010 & \textendash0.70 & $\pm$ 0.05 \\ 
$ \unit{2.5}{\GeV} < $ & \ptlead  & $\leq \unit{5.0}{\GeV}$& 0.514 & $\pm$ 0.005 & 0.087 & $\pm$ 0.012 & \textendash0.89 & $\pm$ 0.05 \\ 
$ \unit{5.0}{\GeV} < $ & \ptlead  & $ \leq \unit{7.5}{\GeV}$  & 0.490 & $\pm$ 0.006 & 0.099 & $\pm$ 0.010 & \textendash0.76 & $\pm$ 0.05 \\ 
$ \unit{7.5}{\GeV} < $ & \ptlead  & $\leq \unit{10}{\GeV}$ & 0.459 & $\pm$ 0.007 & 0.107 & $\pm$ 0.009 & \textendash0.54 & $\pm$ 0.08 \\ 
& \ptlead  & $> \unit{10}{\GeV}$  & 0.415 & $\pm$ 0.010 & 0.117 & $\pm$ 0.011 & \textendash0.28 & $\pm$ 0.13 \\ 
\midrule
&&&\multicolumn{6}{c}{\textbf{Transverse Sphericity}} \\
\multicolumn{3}{c}{\textbf{\ptlead range}} & \multicolumn{2}{l}{\hspace{.25in}Mean} & \multicolumn{2}{l}{\hspace{.25in}RMS} & \multicolumn{2}{c}{Skewness} \\		 	
\midrule
$ \unit{0.5}{\GeV} < $ &  \ptlead & $ \leq \unit{2.5}{\GeV}$ & 0.618 & $\pm$ 0.005 & 0.190 & $\pm$ 0.006 & \textendash0.35 & $\pm$ 0.05 \\ 
$ \unit{2.5}{\GeV} < $ & \ptlead & $ \leq \unit{5.0}{\GeV}$ 	& 0.579 & $\pm$ 0.013 & 0.204 & $\pm$ 0.003 & \textendash 0.28 & $\pm$ 0.12 \\ 
$ \unit{5.0}{\GeV} <  $ & \ptlead &   $\leq  \unit{7.5}{\GeV}$ & 0.449 & $\pm$ 0.019 & 0.206 &$\pm$ 0.002 & 0.16 & $\pm$ 0.24 \\ 
$ \unit{7.5}{\GeV} < $&  \ptlead &  $\leq \unit{10}{\GeV}$ & 0.337 & $\pm$ 0.017 & 0.183 &$\pm$ 0.004 & 0.57 & $\pm$ 0.09 \\ 
& \ptlead  &  $ > \unit{10}{\GeV}$   & 0.230 & $\pm$ 0.024 & 0.157 &$\pm$ 0.007 & 1.06 & $\pm$ 0.04 \\ 

\bottomrule
\end{tabular}
\label{tab:results}
\end{center}
\end{table*}



\section{Conclusions}
\label{sec:conclusions}

The event shape observables, transverse thrust, transverse thrust minor, and
transverse sphericity, 
have been measured in inelastic proton--proton collisions at
$\sqrt{s}=\unit{7}\TeV$ requiring at least six charged particles per event
selected by a minimum-bias trigger.
The distributions and mean values have been 
compared to predictions of different MC models
tuned to inclusive particle distributions and underlying event data.
The dependence of the event shapes 
on the number of charged particles, on the sum of  charged particle \pT\, and on the leading charged particle \pT\ has been studied.

The distributions of all three event shape variables show an evolution toward 
less spherical events as \ptlead
increases, but the effect is smaller for transverse thrust and thrust minor compared to transverse sphericity.
The dependence of the event shape
mean values as functions of \Nchg and \ptsum is similar, 
due the correlation between the two variables~\cite{:2010ir}.
For each variable, the evolution toward a more spherical event shape with increasing multiplicity is rapid initially and slows at higher multiplicities.
All tested MC generators underestimate the fraction of events of spherical character and none reproduces accurately the
event shape distributions. The MC tunes based on the properties of the underlying event show in general better 
agreement with the data than those based on the
inclusive distributions measured in minimum-bias events.
The \pythiasix MC generator with the Z1 tune provides the most accurate description
of the observed distributions 
presented in this analysis,
but the level of agreement
is still not satisfactory over the whole range of the data.
These measurements provide information complementary to inclusive particle distributions and thus they are useful for improving the
MC description of inelastic proton--proton collisions at the LHC.

\FloatBarrier


\section{Acknowledgements}


We thank CERN for the very successful operation of the LHC, as well as the
support staff from our institutions without whom ATLAS could not be
operated efficiently.

We acknowledge the support of ANPCyT, Argentina; YerPhI, Armenia; ARC,
Australia; BMWF, Austria; ANAS, Azerbaijan; SSTC, Belarus; CNPq and FAPESP,
Brazil; NSERC, NRC and CFI, Canada; CERN; CONICYT, Chile; CAS, MOST and NSFC,
China; COLCIENCIAS, Colombia; MSMT CR, MPO CR and VSC CR, Czech Republic;
DNRF, DNSRC and Lundbeck Foundation, Denmark; EPLANET and ERC, European Union;
IN2P3-CNRS, CEA-DSM/IRFU, France; GNAS, Georgia; BMBF, DFG, HGF, MPG and AvH
Foundation, Germany; GSRT, Greece; ISF, MINERVA, GIF, DIP and Benoziyo Center,
Israel; INFN, Italy; MEXT and JSPS, Japan; CNRST, Morocco; FOM and NWO,
Netherlands; RCN, Norway; MNiSW, Poland; GRICES and FCT, Portugal; MERYS
(MECTS), Romania; MES of Russia and ROSATOM, Russian Federation; JINR; MSTD,
Serbia; MSSR, Slovakia; ARRS and MVZT, Slovenia; DST/NRF, South Africa;
MICINN, Spain; SRC and Wallenberg Foundation, Sweden; SER, SNSF and Cantons of
Bern and Geneva, Switzerland; NSC, Taiwan; TAEK, Turkey; STFC, the Royal
Society and Leverhulme Trust, United Kingdom; DOE and NSF, United States of
America.

The crucial computing support from all WLCG partners is acknowledged
gratefully, in particular from CERN and the ATLAS Tier-1 facilities at
TRIUMF (Canada), NDGF (Denmark, Norway, Sweden), CC-IN2P3 (France),
KIT/GridKA (Germany), INFN-CNAF (Italy), NL-T1 (Netherlands), PIC (Spain),
ASGC (Taiwan), RAL (UK) and BNL (USA) and in the Tier-2 facilities
worldwide.


\bibliographystyle{apsrev4-1}

\providecommand{\href}[2]{#2}\begingroup\raggedright\endgroup

\clearpage 
\begin{flushleft}
{\Large The ATLAS Collaboration}

\bigskip

G.~Aad$^{\rm 48}$,
T.~Abajyan$^{\rm 21}$,
B.~Abbott$^{\rm 111}$,
J.~Abdallah$^{\rm 12}$,
S.~Abdel~Khalek$^{\rm 115}$,
A.A.~Abdelalim$^{\rm 49}$,
O.~Abdinov$^{\rm 11}$,
R.~Aben$^{\rm 105}$,
B.~Abi$^{\rm 112}$,
M.~Abolins$^{\rm 88}$
% ...
\bigskip
\\
$^{1}$ School of Chemistry and Physics, University of Adelaide, North Terrace Campus, 5000, SA, Australia\\
% ...
$^{11}$ Institute of Physics, Azerbaijan Academy of Sciences, Baku, Azerbaijan\\
$^{12}$ Institut de F{\'\i}sica d'Altes Energies and Departament de F{\'\i}sica de la Universitat Aut{\`o}noma de Barcelona and ICREA, Barcelona, Spain\\
% ...
$^{21}$ Physikalisches Institut, University of Bonn, Bonn, Germany\\
% ...
$^{48}$ Fakult{\"a}t f{\"u}r Mathematik und Physik, Albert-Ludwigs-Universit{\"a}t, Freiburg, Germany\\
$^{49}$ Section de Physique, Universit{\'e} de Gen{\`e}ve, Geneva, Switzerland\\
% ...
$^{88}$ Department of Physics and Astronomy, Michigan State University, East Lansing MI, United States of America\\
% ...
$^{105}$ Nikhef National Institute for Subatomic Physics and University of Amsterdam, Amsterdam, Netherlands\\
% ...
$^{111}$ Homer L. Dodge Department of Physics and Astronomy, University of Oklahoma, Norman OK, United States of America\\
$^{112}$ Department of Physics, Oklahoma State University, Stillwater OK, United States of America\\
% ...
$^{115}$ LAL, Universit{\'e} Paris-Sud and CNRS/IN2P3, Orsay, France\\
% ...
\end{flushleft}


\end{document}